\documentclass[twocolumn,aps,prl,superscriptaddress]{revtex4-2}
\usepackage[utf8]{inputenc}
\usepackage{amsmath}
\usepackage{stix2}
\usepackage{bm}
\usepackage{xcolor}
\usepackage{graphicx}
\usepackage{outlines}

\begin{document}

\title{Orbital Magnetization of Correlated States in Twisted Bilayer Transition Metal Dichalcogenides}

\author{Xiaoyu Liu}
\affiliation{Department of Materials Science and Engineering, University of Washington, Seattle, WA 98195, USA}
\author{Chong Wang}
\affiliation{Department of Materials Science and Engineering, University of Washington, Seattle, WA 98195, USA}
\author{Haoran Chen}
\affiliation{Department of Materials Science and Engineering, University of Washington, Seattle, WA 98195, USA}
\author{Xiao-Wei Zhang}
\affiliation{Department of Materials Science and Engineering, University of Washington, Seattle, WA 98195, USA}
\author{Ting Cao}
\email{tingcao@uw.edu}
\affiliation{Department of Materials Science and Engineering, University of Washington, Seattle, WA 98195, USA}
\author{Di Xiao}
\email{dixiao@uw.edu}
\affiliation{Department of Materials Science and Engineering, University of Washington, Seattle, WA 98195, USA}
\affiliation{Department of Physics, University of Washington, Seattle, WA 98195, USA}
\affiliation{Pacific Northwest National Laboratory, Richland, WA, USA}

\begin{abstract}
Recent observations of quantum anomalous Hall effects in moir\'e systems have revealed the emergence of interaction-driven ferromagnetism with significant orbital contributions. 
To capture this physics, we extend the modern theory of orbital magnetization to Hartree–Fock states and show that the standard expression remains valid with Hartree–Fock orbitals and Hamiltonians. 
We then benchmark our theory against the extended Kane–Mele–Hubbard model in a weak field, which yields excellent agreement with direct numerical calculations.
Applying our theory to twisted MoTe$_2$ bilayers, we find orbital magnetization of order one Bohr magneton per moir{\'e} cell with a non-monotonic twist-angle dependence. Our work establishes a general theory of orbital magnetization in interacting moir{\'e} systems and provides quantitative guidance for interpreting recent experiments.
\end{abstract}

\maketitle

\textit{Introduction}—Moir{\'e} materials have emerged as a fertile ground for correlated and topological phases. 
Spontaneous time-reversal symmetry breaking and interaction-driven ferromagnetism have been observed in systems ranging from graphene-based moir{\'e} superlattices to twisted bilayer transition-metal dichalcogenides (tTMDs)~\cite{sharpe2019emergent,serlin2020intrinsic,chen2020tunable,polshyn2020electrical,tschirhart2021imaging,anderson2023programming,cai2023signatures,park2023observation,lu2024fractional,redekop2024direct,lu2025extended,park2025ferromagnetism,li2025universal}. Under appropriate conditions, these systems host both integer and fractional quantum anomalous Hall effects ~\cite{serlin2020intrinsic,li2021quantum,cai2023signatures,park2023observation,zeng2023thermodynamic,xu2023observation,foutty2024mapping,lu2024fractional,lu2025extended}, making them ideal platforms to study interaction-driven magnetism tied to band topology.

Unlike conventional ferromagnets, where spin exchange dictates the order, magnetism in moir{\'e} systems is driven primarily by valley polarization and carries substantial orbital contributions. 
Consequently, the net magnetization, which determines valley occupancy and the slope of Landau-fan diagrams in an external magnetic field, reflects a competition between spin and orbital contributions.
In twisted bilayer graphene, where spin moments are quenched, nearly symmetric fan diagrams are attributed to opposite signs of orbital magnetization on the $p$- and $n$-doped sides of the gap~\cite{polshyn2020electrical,zhu2020voltage,wang2024phase}. In contrast, tTMDs exhibit pronounced asymmetry in the fan diagrams: only a negative slope branch in tMoTe$_2$ and a positive slope branch in tWSe$_2$ have been reported at filling factor $\nu = -1$~\cite{cai2023signatures,park2023observation,foutty2024mapping}. These striking differences have been linked to changes in band topology~\cite{zhang2024polarization,wang2024fractional,reddy2023fractional}, but a quantitative treatment of orbital magnetization in correlated states has been lacking.

From a theoretical standpoint, orbital magnetization in periodic systems is subtle because the position operator is ill-defined.
The modern theory resolves this by recasting magnetization in terms of Berry-phase quantities~\cite{Xiao_Berry_2005,Shi_Quantum_2007,Thonhauser_Orbital_2005,Ceresoli_Orbital_2006},
\begin{equation}\label{eq:orbmom}
\bm{M} = -\frac{i e}{2 \hbar V}  \sum_{n\bm{k}}f_{n\bm{k}}
\bigg\langle\frac{\partial u_{n\bm{k}}}{\partial\bm{k}}\bigg|
\times[\hat{H}_0(\bm{k})+\epsilon_n(\bm{k})-2\mu]
\bigg|\frac{\partial u_{n\bm{k}}}{\partial\bm{k}}\bigg\rangle,
\end{equation}
where $u_{n\bm{k}}$ and $\epsilon_{n\bm{k}}$ are Bloch functions and energies of the non-interacting Hamiltonian $\hat{H}_0$, $V$ is the volume of the system, and $f_{n\bm{k}}$ is the occupation number. This formula links magnetization to Berry curvature and yields the St\v{r}eda formula $\partial M/\partial\mu = Ce/h$ in an insulator, where $C$ is the Chern number. 
Importantly, unlike spin magnetization, Eq.~\eqref{eq:orbmom} carries no explicit dependence on the moir\'e cell size, implying that orbital contributions should become increasingly prominent as the twist angle is reduced and spin contributions diluted. The outstanding question is whether this framework remains valid in the presence of strong interactions.  Indeed, although moir\'e flat bands strongly enhance correlations, it has often been assumed, without proof, that Eq.~\eqref{eq:orbmom} holds with Hartree–Fock orbitals~\cite{fan2024orbital,xie2025unconventional,zhu2020voltage}.  Establishing this rigorously is crucial for disentangling spin and orbital contributions in ongoing experiments.

In this work, we show that the modern theory of orbital magnetization remains valid for Hartree-Fock states in the static limit, provided Eq.~\eqref{eq:orbmom} is evaluated with the Hartree-Fock orbitals and Hamiltonian. 
We benchmark this result against total-energy calculations of the Kane–Mele–Hubbard model with magnetic field incorporated via the Hofstadter spectrum, finding excellent agreement.
Applying our theory to twisted MoTe$_2$ bilayers with twist angles $1.89^\circ$–$3.89^\circ$ using Wannier functions constructed from \textit{ab initio} bands, we obtain orbital magnetization of order one Bohr magneton per moir\'e cell.
It is non-monotonic in the twist angle and comparable in size to spin contributions, yet follows distinct trends dictated by band topology.
These findings establish a consistent framework for including interactions in orbital magnetization at the Hartree-Fock level, clarifying their role in moir\'e ferromagnets and providing a firm basis for interpreting Landau-fan asymmetry and tunable quantum anomalous Hall effects.

\textit{Orbital magnetization in Hartree-Fock calculations}.---The orbital magnetization is defined as the response of the grand potential to an external magnetic field, introduced via minimal coupling. 
We therefore begin by examining how a Hartree–Fock system responds to a perturbation. The Hamiltonian of an interacting system can be written as $\hat{H} = \hat{H}_0 + \hat{H}_{\rm int}$, where $\hat{H}_0$ denotes the non-interacting Hamiltonian and $\hat{H}_\text{int}$ accounts for the Coulomb interaction. 
Within the Hartree-Fock approximation, the four-fermion operator $\hat{H}_\text{int}$ is reduced to an effective two-fermion operator $\hat{H}_\text{MF}[\rho]$, expressed as a functional of the one-body reduced density matrix $\rho$. 
We denote $\hat{H}_\text{HF}=\hat{H}_0+\hat{H}_\text{MF}$ as the Hartree-Fock Hamiltonian. 
Solving the Hartree-Fock problem requires a self-consistency loop, and the quasi-orbitals obtained from this procedure are referred to as Hartree–Fock orbitals.

Now suppose that the system is subject to a perturbation $\hat{V}_{\rm ext}$.
According to standard perturbation theory, 
\begin{equation}\label{eq:delta_phi1}
|\delta\phi_\alpha^{(1)}\rangle=\sum_{\beta\neq\alpha}\frac{|\phi_\beta\rangle \langle \phi_\beta| \hat{V}_{\rm ext} | \phi_\alpha \rangle}{\epsilon_\alpha-\epsilon_\beta}.
\end{equation}
Here $\phi_\alpha$ is the quasi-orbital of the Hartree-Fock system, and $\epsilon_\alpha$ is the quasi-energy of this orbital. In a crystal, $\alpha$ includes both the band index $n$ and the $\bm{k}$-index.

In the meantime, the variation of Hartree-Fock orbitals will modify the one-body reduced density matrix $\delta\rho$, which in turn induces a change in the Hartree-Fock Hamiltonian $\delta\hat{H}_\text{HF}[\rho]$. 
This yields an additional first-order correction to the quasi-orbital:
\begin{equation}\label{eq:delta_phi2}
|\delta\phi_\alpha^{(2)}\rangle=\sum_{\beta\neq\alpha}\frac{|\phi_\beta\rangle \langle \phi_\beta| \delta\hat{H}_\mathrm{HF} | \phi_\alpha \rangle}{\epsilon_\alpha-\epsilon_\beta}.
\end{equation}
Therefore, an initial update in quasi-orbitals due to $\hat V_\text{ext}$ changes the Hartree-Fock Hamiltonian $\hat H_\text{HF}$, which then feeds back into the orbitals. 
The Hartree-Fock response must therefore be obtained self-consistently by iterating these updates until the calculation converges. 
This process is similar to the density functional perturbation theory. 

Self-consistent perturbative schemes are computationally costly and can make the physical response harder to interpret. 
Here, in the case of orbital magnetization, we show that \textit{such self-consistency is unnecessary}: Eq.~\eqref{eq:orbmom} can be evaluated directly by substituting the Hartree–Fock Hamiltonian $\hat{H}_{\rm HF}(\bm{k})$ for the non-interacting Hamiltonian and the Hartree–Fock orbitals for the non-interacting Bloch states.

Our proof is built on the quantum mechanical derivation of Eq.~\eqref{eq:orbmom} in Ref.~[\onlinecite{Shi_Quantum_2007}].
To avoid the difficulties in handling a uniform magnetic field, a periodic field is introduced as $\bm{B} (\bm{r}) = B \cos (\bm{q} \cdot \bm{r}) \bm{e}_z$, with the vector potential $\bm{A} (\bm{r}) = - B \sin (\bm{q} \cdot \bm{r}) \bm{e}_x / q$ and $\bm{e}_z$ the unit vector in the $+z$ direction. $\bm{A}(\bm{r})$ couples to the system as $\hat{V}_B = e[\hat{\bm{v}}_0\cdot\bm{A}(\bm{r})+\bm{A}(\bm{r})\cdot\hat{\bm{v}}_0]/2$, where $\hat{\bm{v}}_0 = -i [\hat{\bm{r}}, \hat{H}] / \hbar = -i [\hat{\bm{r}}, \hat{H}_0] / \hbar$ is the bare velocity operator, which is not modified by the Coulomb interaction $\hat{H}_\text{int}$.

From the definition of orbital magnetization, in the limit of $q \to 0$,
\begin{equation}
\delta K(\bm{r})=-\bm{M}\cdot \bm{B}(\bm{r}),
\label{eq:KvsB}
\end{equation}
where $K = E - \mu N$ is the grand potential at zero temperature, $E$ is the energy of the system, and $N$ is the particle number. $K(\bm{r})$ is the local density of the grand potential, which can be evaluated as
\begin{equation}
    \begin{aligned}
        \delta K(\bm{r})= & \frac{1}{2} \left[ \sum_{\alpha} f_{\alpha}\langle\phi_{\alpha} | \{(\hat{H}_{\rm HF} - \mu\hat{N}),\delta(\hat{\bm{r}}-\bm{r})\} | \delta \phi_{\alpha}\rangle + c.c. \right] \\
        & + \frac{1}{2} \sum_{n\bm{k}} f_{\alpha} \langle \phi_{\alpha}| \{\hat{V}_B , \delta(\hat{\bm{r}}-\bm{r})\}|\phi_{\alpha}\rangle.
    \end{aligned}
    \label{eq:local_K}
\end{equation}
Orbital magnetization $\bm{M}$ can therefore be extracted by comparing Eq.~\eqref{eq:KvsB}
and Eq.~\eqref{eq:local_K}.  The last term in Eq.~\eqref{eq:local_K} does not contribute to the magnetization~\cite{Shi_Quantum_2007}.  

Let us first consider the contribution from$|\delta \phi_\alpha^{(1)}\rangle$, which is given by (see the Supplemental Material~\cite{supp} for details)
\begin{equation}
\begin{aligned}
    M_z^{(1)} =& \lim_{q\to 0}\frac{e}{4 q V}\textrm{Im}\sum_{nn'\bm{k}}[
    (\epsilon_{n\bm{k}}+\epsilon_{n'\bm{k}+\bm{q}}-2\mu) 
    \langle u_{n\bm{k}}| u_{n'\bm{k}+\bm{q}} \rangle \\
    &\langle u_{n'\bm{k}+\bm{q}}| \hat{v}_{0,x}(\bm{k}+\bm{q}) +  \hat{v}_{0,x}(\bm{k})|u_{n\bm{k}}\rangle
     \frac{f_{n\bm{k}}-f_{n'\bm{k}+\bm{q}}}{\epsilon_{n\bm{k}}-\epsilon_{n'\bm{k}+\bm{q}}}],
\end{aligned}
\label{eq:Mz1}
\end{equation}
where $\hat{v}_{0,x}(\bm{k}) = e^{-i \bm{k} \cdot \bm{r}} \hat{v}_{0,x} e^{i \bm{k} \cdot \bm{r}}$. We note that Eq.~\eqref{eq:Mz1} refers directly to the bare velocity operator $\hat{\bm{v}}_0$ and does not satisfy the St\v{r}eda formula.  This indicates that $M_z^{(1)}$ is incomplete by itself.

The contribution of $|\delta\phi_{n\bm{k}}^{(2)}\rangle$ to orbital magnetization is denoted as $M_z^{(2)}$. We note that $\delta \hat{H}_{\rm HF}$ is entirely contributed by the change of $\hat{H}_{\rm MF}$. Generally, for an external perturbation, $\delta \hat{H}_{\rm HF}$ needs to be obtained in a self-consistent fashion. However, since eventually the limit $q \to 0$ will be taken, we only need to analyze the behavior of $\delta \hat{H}_{\rm HF}$ in the small $q$ limit where $\bm{A}$ varies slowly in real space. For a constant $\bm{A}$ (a pure gauge transformation), it can be shown that $\delta \hat{H}_{\rm HF} = e\hat{\bm{v}}_{\rm MF} \cdot \bm{A}$, where $\hat{\bm{v}}_{\rm MF} = -i [\hat{\bm{r}}, \hat{H}_{\rm MF}] / \hbar$.
Assuming locality in Hartree-Fock calculations, for a slowly varying vector potential $\bm{A}(\bm{r})$, $\delta \hat{H}_{\rm HF} \approx e[\hat{\bm{v}}_{\rm MF} \cdot \bm{A}(\hat{\bm{r}}) + \bm{A}(\hat{\bm{r}}) \cdot \hat{\bm{v}}_{\rm MF}] / 2 $. Here, we have symmetrized $\hat{\bm{v}}_{\rm MF}$ and $\bm{A}(\hat{\bm{r}})$ to ensure that $\delta \hat{H}_{\rm HF}$ is Hermitian. Other symmetrization schemes are possible, but will not affect the final result. With the knowledge of $\delta \hat{H}_{\rm HF}$, it is straightforward to show that $M_z^{(2)}$ has the same form as $M_z^{(1)}$ in Eq.~\eqref{eq:Mz1}, but with $\hat{\bm{v}}_0$ replaced by $\hat{\bm{v}}_{\rm MF}$.

Putting $M_z^{(1)}$ and $M_z^{(2)}$ together, we find that the relevant velocity operator is $\hat{\bm{v}}_{\rm HF} = \hat{\bm{v}}_0 + \hat{\bm{v}}_{\rm MF} = -i [\hat{\bm{r}},\hat{H}_{\rm HF}] / \hbar$. After some straightforward algebra, the final expression for orbital magnetization is 
\begin{equation}\label{eq:orbmom_hf}
\bm{M} = -\frac{i e}{2\hbar V} \sum_{n\bm{k}}f_{n\bm{k}}
\bigg\langle\frac{\partial u_{n\bm{k}}}{\partial\bm{k}}\bigg|
\times[\hat{H}_\mathrm{HF}(\bm{k})+\epsilon_{n\bm{k}}-2\mu]
\bigg|\frac{\partial u_{n\bm{k}}}{\partial\bm{k}}\bigg\rangle.
\end{equation}
Here, $u_{n\bm{k}}$ and $\epsilon_{n\bm{k}}$ are the periodic part of the Hartree-Fock quasi-orbitals and the quasi-energies, respectively. In other words, orbital magnetization in the Hartree-Fock approximation can be evaluated with the Hartree-Fock Hamiltonian and orbitals as if the system has no interactions.  The St\v{r}eda formula also holds for Eq.~\eqref{eq:orbmom_hf}.

We note that $\delta \hat{H}_{\rm HF} \approx e[\hat{\bm{v}}_{\rm MF} \cdot \bm{A}(\hat{\bm{r}}) + \bm{A}(\hat{\bm{r}}) \cdot \hat{\bm{v}}_{\rm MF}] / 2 $ can be viewed as an expansion of $\delta \hat{H}_{\rm HF}$ in $q$ around $q = 0$. $\bm{A}(\hat{\bm{r}})$ is proportional to $1 / q$. The next-order contribution to $\delta \hat{H}_{\rm HF}$ is expected to be proportional to gradients of $\bm{A}$ and is ${\rm O}(1)$. However, in taking the limit of $q \to 0$, a $1/q$ term has to be paired with $\langle u_{n\bm{k}}|u_{n'\bm{k}+\bm{q}} \rangle$ in Eq.~\eqref{eq:Mz1} in the L'H\^{o}pital's rule. Therefore, an ${\rm O}(1)$ contribution to $\delta \hat{H}_{\rm HF}$ does not contribute to the expression of orbital magnetization. 

The derivation shows that the bare velocity operator $\hat{\bm{v}}_0$ is corrected by the interaction in the context of the Hartree-Fock approximation. This correction is important to keep the mean-field theory gauge invariant. In the language of field theory, the velocity operator corresponds to a vertex in Feynman diagrams that couples to vector potentials. Using the Ward identity and considering the limit of $q \to 0$ (long wavelength) and $\omega \to 0$ (static), we have $\hat{\bm{v}} = -\partial G^{-1}/\partial(\hbar\bm{k})$. Given that $G$ is the Green's function corresponding to $\hat{H}_\mathrm{HF}$, one finds that the bare vertex $\hat{\bm{v}}_0$ should be dressed to generate $\hat{\bm{v}}_\mathrm{HF}$.


In Ref.~[\onlinecite{Shi_Quantum_2007}], it is shown that the non-interacting expression for orbital magnetization remains valid in current and spin density functional theory, where the electron-electron interaction is described by Hartree and exchange-correlation terms. The present derivation demonstrates that the same formal structure of the orbital magnetization expression is preserved within the self-consistent Hartree–Fock framework, despite the presence of nonlocal Fock terms.



\textit{Kane-Mele-Hubbard model}---To test the validity of Eq.~\eqref{eq:orbmom_hf}, we calculate the orbital magnetization of the extended Kane-Mele-Hubbard (KMH) model~\cite{kane2005quantum,kane2005z} on a honeycomb lattice: 
\begin{equation}\label{eq:kmh}
\begin{aligned}
    \hat{H}^{\rm KMH} = & \sum_{\langle i, j\rangle, \sigma} t_1 \hat{c}_{i\sigma}^\dagger \hat{c}_{j\sigma} + 
    \sum_{\langle\langle i, j \rangle\rangle, \sigma} t_2 ~ e^{i\sigma\nu_{ij}\theta} \hat{c}_{i\sigma}^\dagger \hat{c}_{j\sigma} \\
    & +\sum_i U \hat{n}_{i\uparrow} \hat{n}_{i\downarrow} + \sum_{\langle i,j\rangle, \sigma}V \hat{n}_{i\sigma} \hat{n}_{j\sigma'}.
\end{aligned}
\end{equation}
The kinetic part includes the nearest-neighbor hopping $t_1$ and the next-nearest-neighbor hopping $t_2$. $\sigma=\pm1$ for spin up (down), $\nu_{ij}=\pm1$ for clockwise (anticlockwise) next-nearest-neighbor hopping. $\theta$ is the phase of the next-nearest-neighbor hopping and is chosen as $\pi/3$ to represent tTMD~\cite{liu2024gate}. The interacting part includes both the onsite Coulomb interaction $U$ and the nearest-neighbor density-density interaction $V$. In the following discussion, we choose $t_1=1$, $t_2=1/3, U=10$. As shown in previous works~\cite{wu2019topological,anderson2023programming,cai2023signatures,park2023observation,qiu2023interaction,liu2024gate,fan2024orbital}, tTMDs can be modeled by the Kane-Mele-Hubbard model with spins in the model corresponding to the valley degree of freedom.

\begin{figure}
\centering
\includegraphics[width=\columnwidth]{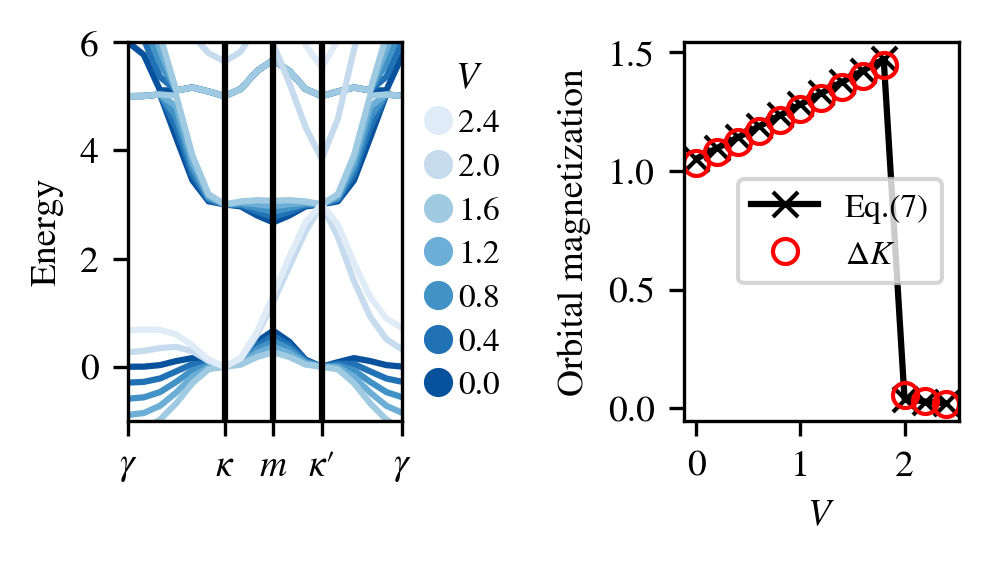}
\caption{(a) Band structure of KMH model with different next-nearest-neighbor Coulomb interaction $V$. (b) Orbital magnetization calculated from Eq.~\eqref{eq:orbmom_hf} (black crosses) and from the finite-difference method (red open dots). \label{fig:kmh}}
\end{figure}

At filling $\nu=-1$, the system is spin(valley)-polarized and breaks the time-reversal symmetry, generating finite orbital magnetization.
With the increase of the next-nearest-neighbor interaction $V$, it undergoes a topological phase transition from $|C|=1$ to $C=0$, as shown in Fig.~\ref{fig:kmh}(a).
This corresponds to a phase transition from a quantum anomalous Hall state to a trivial charge density wave phase.
More discussions are provided in the Supplemental Material~\cite{supp}. 
Before the phase transition, the gap of the system grows linearly with $V$.
The orbital magnetization, calculated using Eq.~\eqref{eq:orbmom_hf} at the valence band maximum (VBM) [black crosses in Fig.~\ref{fig:kmh}(b)] also increases linearly. 
After the phase transition point around $V = 1.8$, it drops abruptly to nearly zero. 
It is worth noting that, unlike in non-interacting systems where orbital magnetization can be evaluated at arbitrary chemical potentials --- in the Hartree-Fock framework it is well-defined only at specific fillings, due to its dependence on the occupation. The value at the conduction band minimum (CBM) can be inferred from the value at VBM and the St\v{r}eda relation $\Delta M/\Delta\mu=Ce/h$.  Accordingly, we report orbital magnetization only at the VBM.

For a direct check, we also calculate the orbital magnetization via its definition $\bm M = -\partial K(\bm B)/\partial \bm B$ numerically.  Here, a weak magnetic field is introduced through the Peierls substitution $t_{ij}\rightarrow t_{ij} \exp [i(e/\hbar)\int_{\bm{r}_i}^{\bm{r}_j} \bm{A}(\bm{r}) d\bm{r}]$, where $\bm{A}$ is the vector potential associated with the magnetic field $\bm{B}$.  We then perform Hartree-Fock calculations of the total energy on top of the resulting Hofstadter spectrum.
To reach small magnetic fields, we need to increase the size of the magnetic unit cell. The convergence test with respect to the size of the magnetic unit cell can be found in Supplemental Material~\cite{supp}.
An important subtlety is that applying a magnetic field to topological systems can alter the density of states~\cite{Xiao_Berry_2005} and change the occupation number for fixed chemical potential $\mu$.  
Therefore, in the Hartree-Fock calculations, the particle number needs to be varied to find the minimal grand potential $K$ for a fixed $\mu$. 
We then extract the orbital magnetization by finite differences.
As shown in Fig.~\ref{fig:kmh}(b), the finite-difference results (red circles) track Eq.~\eqref{eq:orbmom_hf} (black crosses) quantitatively on both sides of the transition, providing strong evidence for the validity of Eq.~\eqref{eq:orbmom_hf}.

\textit{Twist-angle-dependent orbital magnetization moir\'e materials}---Having established the validity of Eq.~\eqref{eq:orbmom_hf}, we now examine the size and sign of the orbital magnetization, and its competition with spin magnetization in moir\'e materials where correlations are essential. Spin magnetization scales as $\mu_B/A_\mathrm{uc}$, with $A_\mathrm{uc}$ the unit-cell area. Since moir\'e superlattices already possess unit cells orders of magnitude larger than atomic crystals, the spin contribution is strongly diluted. In contrast, orbital magnetization -- set by band topology and Berry curvature -- does not explicitly scale with unit-cell size.  Naively, this disparity suggests that orbital effects may dominate in moir\'e systems, particularly when the twist angle is small and the moir\'e unit cell is large.

To test this scenario, we apply Eq.~\eqref{eq:orbmom_hf} to tMoTe$_2$.  Rather than continuum models~\cite{wu2019topological,chatterjee2020symmetry,wang2024fractional,reddy2023fractional,zhang2024polarization,jia2024moire,zhang2025twist,zhang2024universal}, here we use Wannier models constructed from \textit{ab initio} calculations, which realistically incorporate lattice relaxation. 
Following the procedure in Ref.~\cite{wang2025higher}, we build Wannier models at eight different twist angles between $1.89^\circ$ to $3.89^\circ$, and perform Hartree–Fock calculations at filling factor $\nu=-1$.  
We have chosen the dielectric constant $\epsilon=40$ to match the experimentally observed gap~\cite{park2023observation,redekop2024direct}.

\begin{figure}
\centering
\includegraphics[width=\columnwidth]{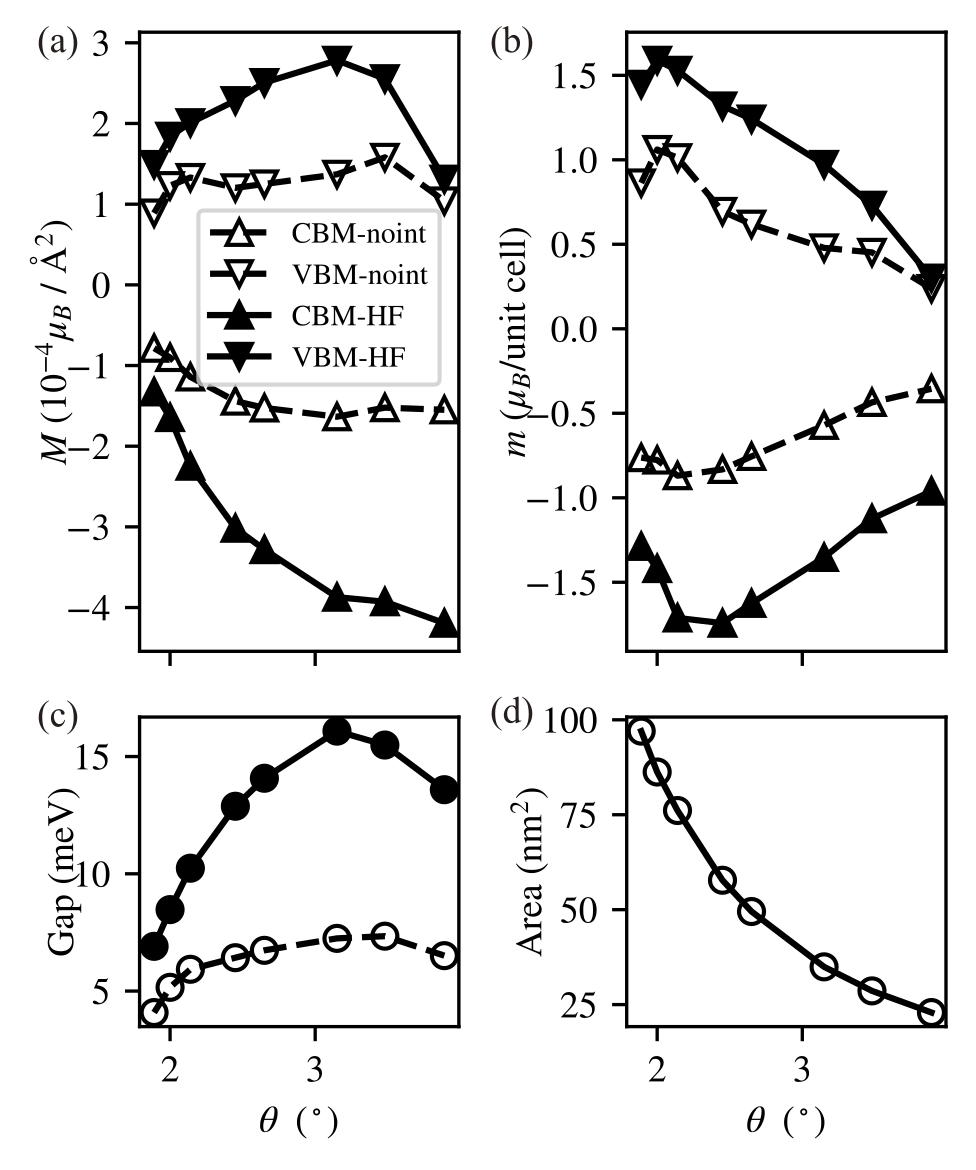}
\caption{(a) Orbital magnetization (b) orbital moment per moir\'e unit cell for tMoTe$_2$. Open symbols and filled symbols represent non-interacting results and Hartree-Fock results, respectively. Upward and downward triangles represent CBM and VBM, respectively. (c) Energy gap between the first and second moir\'e valence bands at various twist angles from Wannier models. Open symbols represent results without interaction, while filled symbols include interaction. (d) Moir\'e unit cell area as a function of the twist angle. \label{fig:wannier}}
\end{figure}

Figure~\ref{fig:wannier}(a) shows the orbital magnetization of tMoTe$_2$ in the $K$ valley versus twist angle for interacting (Hartree–Fock; filled symbols) and non-interacting (open symbols) calculations when the chemical potential is at the VBM (downward triangles) and the CBM (upward triangles). To compare with spin magnetization, we rescale $m$ in Fig.~\ref{fig:wannier}(a) by the moir\'e unit-cell area in Fig.~\ref{fig:wannier}(b). For definitiveness, we discuss CBM in the following. Notably, $m$ at CBM is non-monotonic in $\theta$: it increases as the moir\'e unit cell grows, then decreases for $\theta \lesssim 2.14^\circ$. This behavior arises from competing effects. Decreasing $\theta$ enlarges the unit cell, which tends to enhance orbital magnetization, but it also suppresses inter-cell electronic hopping as the electrons localize -- evidenced by the reduced gap between the first moir\'e valence band and the second band in Fig.~\ref{fig:wannier}(c). Because inter-cell hopping underlies the circulating currents that generate orbital magnetization, its suppression at small $\theta$ reduces $m$. The balance of these trends yields the observed non-monotonic dependence.

The peak value of $m$ at CBM occurs at twist angles between $2.14^\circ$ and $2.45^\circ$, reaching about $1.7\mu_B$ per moir\'e unit cell.
This magnitude is already comparable to a spin moment and thus represents a significant enhancement compared with conventional materials. 
However, in TMDs the total magnetization also includes a substantial atomic orbital contribution from local $d$ orbitals, encoded in the effective $g$ factor in the absence of the moir{\'e} potential.
Experiments and theory indicate that combined spin and atomic orbital moments in monolayer tTMDs can reach $\sim 6$–$8\,\mu_B$ ~\cite{Deilmann2020ab,wozniak2020exciton,robert2021measurement,redekop2024direct}. 
Thus, although orbital magnetization is strongly enhanced and non-negligible, spin plus atomic orbital contributions remain dominant across the twist-angle range studied, consistent with recent nSOT measurements of magnetic fringe fields~\cite{redekop2024direct}.

In summary, we show that the non-interacting equation for orbital magnetization remains valid for Hartree-Fock states, provided that Hartree-Fock orbitals and Hamiltonian are used.  
In realistic systems like tMoTe$_2$, constructing Wannier models on density functional theory calculations -- where structural relaxations and polarizations are taken into consideration -- reveals a complex angle dependence of orbital magnetization. Notably, contrary to the conventional expectation that a smaller twist angle corresponds to stronger interactions, orbital magnetization in fact decreases with decreasing the twist angle. While the large unit cell of moir\'e systems facilitates an enhanced orbital magnetization, across the twist angles considered for tMoTe$_2$, spin and atomic orbital magnetization dominate over orbital magnetization.

\begin{acknowledgments}
We thank Xiaodong Hu, Jian Kang, Junren Shi, Xiaodong Xu, Kaijie Yang, Oskar Vafek for stimulating discussions.  This work is mainly supported by the Center on Programmable Quantum Materials, an Energy Frontier Research Center funded by DOE BES under award DE-SC0019443. The development of advanced codes for moir\'e systems was supported by the Computational Materials Sciences Program funded by the U.S. Department of Energy, Office of Science, Basic Energy Sciences, Materials Sciences, and Engineering Division, PNNL FWP 83557. This research used resources of the National Energy Research Scientific Computing Center, a DOE Office of Science User Facility supported by the Office of Science of the U.S. Department of Energy under Contract No. DE-AC02-05CH11231 using NERSC award BES-ERCAP0032546, BES-ERCAP0033256, and BES-ERCAP0033507. This work was also facilitated through the use of advanced computational, storage, and networking infrastructure provided by the Hyak supercomputer system and funded by the University of Washington Molecular Engineering Materials Center at the University of Washington (NSF MRSEC DMR-2308979). 
\end{acknowledgments}

\textit{Note added.}---We recently became aware of an independent work on similar topics \cite{kang2025orbital}. 

\bibliography{main}

@misc{supp,
    Note = {
    Supplemental Material containing derivation of Eq.(7) and details of both model and moir\'e material calculations. Supplemental Material cites Refs.~\cite{fan2024intrinsic,zhu2024layer,guinea2018electrostatic,zhu2024weak}.
    }
}

@article{Shi_Quantum_2007,
  title = {Quantum Theory of Orbital Magnetization and Its Generalization to Interacting Systems},
  author = {Shi, Junren and Vignale, G. and Xiao, Di and Niu, Qian},
  journal = {Phys. Rev. Lett.},
  volume = {99},
  issue = {19},
  pages = {197202},
  numpages = {4},
  year = {2007},
  month = {Nov},
  publisher = {American Physical Society},
  doi = {10.1103/PhysRevLett.99.197202},
  url = {https://link.aps.org/doi/10.1103/PhysRevLett.99.197202}
}

@article{Xiao_Berry_2005,
  title = {Berry Phase Correction to Electron Density of States in Solids},
  author = {Xiao, Di and Shi, Junren and Niu, Qian},
  journal = {Phys. Rev. Lett.},
  volume = {95},
  issue = {13},
  pages = {137204},
  numpages = {4},
  year = {2005},
  month = {Sep},
  publisher = {American Physical Society},
  doi = {10.1103/PhysRevLett.95.137204},
  url = {https://link.aps.org/doi/10.1103/PhysRevLett.95.137204}
}

@article{Ceresoli_Orbital_2006,
  title = {Orbital magnetization in crystalline solids: Multi-band insulators, Chern insulators, and metals},
  author = {Ceresoli, Davide and Thonhauser, T. and Vanderbilt, David and Resta, R.},
  journal = {Phys. Rev. B},
  volume = {74},
  issue = {2},
  pages = {024408},
  numpages = {13},
  year = {2006},
  month = {Jul},
  publisher = {American Physical Society},
  doi = {10.1103/PhysRevB.74.024408},
  url = {https://link.aps.org/doi/10.1103/PhysRevB.74.024408}
}

@article{Thonhauser_Orbital_2005,
  title = {Orbital Magnetization in Periodic Insulators},
  author = {Thonhauser, T. and Ceresoli, Davide and Vanderbilt, David and Resta, R.},
  journal = {Phys. Rev. Lett.},
  volume = {95},
  issue = {13},
  pages = {137205},
  numpages = {4},
  year = {2005},
  month = {Sep},
  publisher = {American Physical Society},
  doi = {10.1103/PhysRevLett.95.137205},
  url = {https://link.aps.org/doi/10.1103/PhysRevLett.95.137205}
}

@article{wu2019topological,
  title={Topological insulators in twisted transition metal dichalcogenide homobilayers},
  author={Wu, Fengcheng and Lovorn, Timothy and Tutuc, Emanuel and Martin, Ivar and MacDonald, AH},
  journal={Phys. Rev. Lett.},
  volume={122},
  number={8},
  pages={086402},
  year={2019},
  publisher={APS}
}

@article{wang2024fractional,
  title={Fractional Chern insulator in twisted bilayer {MoTe2}},
  author={Wang, Chong and Zhang, Xiao-Wei and Liu, Xiaoyu and He, Yuchi and Xu, Xiaodong and Ran, Ying and Cao, Ting and Xiao, Di},
  journal={Phys. Rev. Lett.},
  volume={132},
  number={3},
  pages={036501},
  year={2024},
  publisher={APS}
}

@article{redekop2024direct,
  title={Direct magnetic imaging of fractional Chern insulators in twisted {MoTe2}},
  author={Redekop, Evgeny and Zhang, Canxun and Park, Heonjoon and Cai, Jiaqi and Anderson, Eric and Sheekey, Owen and Arp, Trevor and Babikyan, Grigory and Salters, Samuel and Watanabe, Kenji and others},
  journal={Nature},
  volume={635},
  number={8039},
  pages={584--589},
  year={2024},
  publisher={Nature Publishing Group UK London}
}

@article{zhang2024polarization,
  title={Polarization-driven band topology evolution in twisted {MoTe$_2$} and {WSe$_2$}},
  author={Zhang, Xiao-Wei and Wang, Chong and Liu, Xiaoyu and Fan, Yueyao and Cao, Ting and Xiao, Di},
  journal={Nat. Commun.},
  volume={15},
  number={1},
  pages={4223},
  year={2024},
  publisher={Nature Publishing Group UK London}
}

@article{jia2024moire,
  title={Moir{\'e} fractional Chern insulators. I. First-principles calculations and continuum models of twisted bilayer {MoTe$_2$}},
  author={Jia, Yujin and Yu, Jiabin and Liu, Jiaxuan and Herzog-Arbeitman, Jonah and Qi, Ziyue and Pi, Hanqi and Regnault, Nicolas and Weng, Hongming and Bernevig, B Andrei and Wu, Quansheng},
  journal={Phys. Rev. B},
  volume={109},
  number={20},
  pages={205121},
  year={2024},
  publisher={APS}
}

@article{wang2025higher,
  title={Higher Landau-Level Analogs and Signatures of Non-Abelian States in Twisted Bilayer MoTe 2},
  author={Wang, Chong and Zhang, Xiao-Wei and Liu, Xiaoyu and Wang, Jie and Cao, Ting and Xiao, Di},
  journal={Phys. Rev. Lett.},
  volume={134},
  number={7},
  pages={076503},
  year={2025},
  publisher={APS}
}

@article{foutty2024mapping,
  title={Mapping twist-tuned multiband topology in bilayer WSe2},
  author={Foutty, Benjamin A and Kometter, Carlos R and Devakul, Trithep and Reddy, Aidan P and Watanabe, Kenji and Taniguchi, Takashi and Fu, Liang and Feldman, Benjamin E},
  journal={Science},
  volume={384},
  number={6693},
  pages={343--347},
  year={2024},
  publisher={American Association for the Advancement of Science}
}

@article{zhu2020voltage,
  title={Voltage-controlled magnetic reversal in orbital Chern insulators},
  author={Zhu, Jihang and Su, Jung-Jung and MacDonald, Allan H},
  journal={Phys. Rev. Lett.},
  volume={125},
  number={22},
  pages={227702},
  year={2020},
  publisher={APS}
}

@article{polshyn2020electrical,
  title={Electrical switching of magnetic order in an orbital Chern insulator},
  author={Polshyn, Hryhoriy and Zhu, Jihang and Kumar, Manish A and Zhang, Yuxuan and Yang, Fangyuan and Tschirhart, Charles L and Serlin, Marec and Watanabe, Kenji and Taniguchi, Takashi and MacDonald, Allan H and others},
  journal={Nature},
  volume={588},
  number={7836},
  pages={66--70},
  year={2020},
  publisher={Nature Publishing Group UK London}
}

@article{liu2024gate,
  title={Gate-tunable antiferromagnetic Chern insulator in twisted bilayer transition metal dichalcogenides},
  author={Liu, Xiaoyu and He, Yuchi and Wang, Chong and Zhang, Xiao-Wei and Cao, Ting and Xiao, Di},
  journal={Phys. Rev. Lett.},
  volume={132},
  number={14},
  pages={146401},
  year={2024},
  publisher={APS}
}

@article{Deilmann2020ab,
  title = {Ab Initio Studies of Exciton $g$ Factors: Monolayer Transition Metal Dichalcogenides in Magnetic Fields},
  author = {Deilmann, Thorsten and Kr\"uger, Peter and Rohlfing, Michael},
  journal = {Phys. Rev. Lett.},
  volume = {124},
  issue = {22},
  pages = {226402},
  numpages = {7},
  year = {2020},
  month = {Jun},
  publisher = {American Physical Society},
  doi = {10.1103/PhysRevLett.124.226402},
  url = {https://link.aps.org/doi/10.1103/PhysRevLett.124.226402}
}

@article{robert2021measurement,
  title = {Measurement of Conduction and Valence Bands $g$-Factors in a Transition Metal Dichalcogenide Monolayer},
  author = {Robert, C. and Dery, H. and Ren, L. and Van Tuan, D. and Courtade, E. and Yang, M. and Urbaszek, B. and Lagarde, D. and Watanabe, K. and Taniguchi, T. and Amand, T. and Marie, X.},
  journal = {Phys. Rev. Lett.},
  volume = {126},
  issue = {6},
  pages = {067403},
  numpages = {9},
  year = {2021},
  month = {Feb},
  publisher = {American Physical Society},
  doi = {10.1103/PhysRevLett.126.067403},
  url = {https://link.aps.org/doi/10.1103/PhysRevLett.126.067403}
}

@article{wozniak2020exciton,
  title = {Exciton $g$ factors of van der Waals heterostructures from first-principles calculations},
  author = {Wo\ifmmode \acute{z}\else \'{z}\fi{}niak, Tomasz and Faria Junior, Paulo E. and Seifert, Gotthard and Chaves, Andrey and Kunstmann, Jens},
  journal = {Phys. Rev. B},
  volume = {101},
  issue = {23},
  pages = {235408},
  numpages = {11},
  year = {2020},
  month = {Jun},
  publisher = {American Physical Society},
  doi = {10.1103/PhysRevB.101.235408},
  url = {https://link.aps.org/doi/10.1103/PhysRevB.101.235408}
}

@article{chatterjee2020symmetry,
  title={Symmetry breaking and skyrmionic transport in twisted bilayer graphene},
  author={Chatterjee, Shubhayu and Bultinck, Nick and Zaletel, Michael P},
  journal={Phys. Rev. B},
  volume={101},
  number={16},
  pages={165141},
  year={2020},
  publisher={APS}
}

@article{kane2005quantum,
  title={Quantum spin {Hall} effect in graphene},
  author={Kane, Charles L and Mele, Eugene J},
  journal={Phys. Rev. Lett.},
  volume={95},
  number={22},
  pages={226801},
  year={2005},
  publisher={APS}
}

@article{kane2005z,
  title={{Z2} topological order and the quantum spin Hall effect},
  author={Kane, Charles L and Mele, Eugene J},
  journal={Phys. Rev. Lett.},
  volume={95},
  number={14},
  pages={146802},
  year={2005},
  publisher={APS}
}

@article{anderson2023programming,
  title={Programming correlated magnetic states with gate-controlled moir{\'e} geometry},
  author={Anderson, Eric and Fan, Feng-Ren and Cai, Jiaqi and Holtzmann, William and Taniguchi, Takashi and Watanabe, Kenji and Xiao, Di and Yao, Wang and Xu, Xiaodong},
  journal={Science},
  volume={381},
  number={6655},
  pages={325--330},
  year={2023},
  publisher={American Association for the Advancement of Science}
}

@article{cai2023signatures,
  title={Signatures of Fractional Quantum Anomalous {Hall} States in Twisted {MoTe$_2$}},
  author={Cai, Jiaqi and Anderson, Eric and Wang, Chong and Zhang, Xiaowei and Liu, Xiaoyu and Holtzmann, William and Zhang, Yinong and Fan, Fengren and Taniguchi, Takashi and Watanabe, Kenji and others},
  journal={Nature},
  pages={1--3},
  year={2023},
  publisher={Nature Publishing Group UK London}
}

@article{park2023observation,
  title={Observation of fractionally quantized anomalous Hall effect},
  author={Park, Heonjoon and Cai, Jiaqi and Anderson, Eric and Zhang, Yinong and Zhu, Jiayi and Liu, Xiaoyu and Wang, Chong and Holtzmann, William and Hu, Chaowei and Liu, Zhaoyu and others},
  journal={Nature},
  volume={622},
  number={7981},
  pages={74--79},
  year={2023},
  publisher={Nature Publishing Group UK London}
}

@article{lu2024fractional,
  title={Fractional quantum anomalous Hall effect in multilayer graphene},
  author={Lu, Zhengguang and Han, Tonghang and Yao, Yuxuan and Reddy, Aidan P and Yang, Jixiang and Seo, Junseok and Watanabe, Kenji and Taniguchi, Takashi and Fu, Liang and Ju, Long},
  journal={Nature},
  volume={626},
  number={8000},
  pages={759--764},
  year={2024},
  publisher={Nature Publishing Group UK London}
}

@article{serlin2020intrinsic,
  title={Intrinsic quantized anomalous Hall effect in a moir{\'e} heterostructure},
  author={Serlin, Marec and Tschirhart, CL and Polshyn, Hryhoriy and Zhang, Yuxuan and Zhu, Jiacheng and Watanabe, Kenji and Taniguchi, Takashi and Balents, L and Young, AF},
  journal={Science},
  volume={367},
  number={6480},
  pages={900--903},
  year={2020},
  publisher={American Association for the Advancement of Science}
}

@article{li2021quantum,
  title={Quantum anomalous Hall effect from intertwined moir{\'e} bands},
  author={Li, Tingxin and Jiang, Shengwei and Shen, Bowen and Zhang, Yang and Li, Lizhong and Tao, Zui and Devakul, Trithep and Watanabe, Kenji and Taniguchi, Takashi and Fu, Liang and others},
  journal={Nature},
  volume={600},
  number={7890},
  pages={641--646},
  year={2021},
  publisher={Nature Publishing Group UK London}
}

@article{xu2023observation,
  title={Observation of integer and fractional quantum anomalous Hall effects in twisted bilayer MoTe 2},
  author={Xu, Fan and Sun, Zheng and Jia, Tongtong and Liu, Chang and Xu, Cheng and Li, Chushan and Gu, Yu and Watanabe, Kenji and Taniguchi, Takashi and Tong, Bingbing and others},
  journal={Phys. Rev. X},
  volume={13},
  number={3},
  pages={031037},
  year={2023},
  publisher={APS}
}

@article{sharpe2019emergent,
  title={Emergent ferromagnetism near three-quarters filling in twisted bilayer graphene},
  author={Sharpe, Aaron L and Fox, Eli J and Barnard, Arthur W and Finney, Joe and Watanabe, Kenji and Taniguchi, Takashi and Kastner, MA and Goldhaber-Gordon, David},
  journal={Science},
  volume={365},
  number={6453},
  pages={605--608},
  year={2019},
  publisher={American Association for the Advancement of Science}
}

@article{reddy2023fractional,
  title={Fractional quantum anomalous Hall states in twisted bilayer {MoTe$_2$} and {WSe$_2$}},
  author={Reddy, Aidan P and Alsallom, Faisal and Zhang, Yang and Devakul, Trithep and Fu, Liang},
  journal={Physical Review B},
  volume={108},
  number={8},
  pages={085117},
  year={2023},
  publisher={APS}
}

@misc{zhang2025twist,
      title={Twist-angle transferable continuum model and second flat Chern band in twisted {MoTe$_2$} and {WSe$_2$}}, 
      author={Xiao-Wei Zhang and Kaijie Yang and Chong Wang and Xiaoyu Liu and Ting Cao and Di Xiao},
      year={2025},
      eprint={2508.17673},
      archivePrefix={arXiv},
      url={https://arxiv.org/abs/2508.17673}, 
}

@misc{zhang2024universal,
      title={Universal Moir\'e-Model-Building Method without Fitting: Application to Twisted {MoTe$_2$} and {WSe$_2$}}, 
      author={Yan Zhang and Hanqi Pi and Jiaxuan Liu and Wangqian Miao and Ziyue Qi and Nicolas Regnault and Hongming Weng and Xi Dai and B. Andrei Bernevig and Quansheng Wu and Jiabin Yu},
      year={2024},
      eprint={2411.08108},
      archivePrefix={arXiv},
      url={https://arxiv.org/abs/2411.08108}, 
}

@article{fan2024orbital,
  title = {Orbital Chern insulator at $\nu=-2$ in twisted {$\mathrm{MoTe}_2$}},
  author = {Fan, Feng-Ren and Xiao, Cong and Yao, Wang},
  journal = {Phys. Rev. B},
  volume = {109},
  issue = {4},
  pages = {L041403},
  numpages = {5},
  year = {2024},
  month = {Jan},
  publisher = {American Physical Society},
  doi = {10.1103/PhysRevB.109.L041403},
  url = {https://link.aps.org/doi/10.1103/PhysRevB.109.L041403}
}

@misc{xie2025unconventional,
      title={Unconventional Orbital Magnetism in Graphene-based Fractional Chern Insulators}, 
      author={Jian Xie and Zaizhe Zhang and Xi Chen and Yves H. Kwan and Zihao Huo and Jonah Herzog-Arbeitman and Liangliang Guo and Kenji Watanabe and Takashi Taniguchi and Kaihui Liu and X. C. Xie and B. Andrei Bernevig and Zhi-Da Song and Xiaobo Lu},
      year={2025},
      eprint={2506.01485},
      archivePrefix={arXiv},
      url={https://arxiv.org/abs/2506.01485}, 
}

@article{park2025ferromagnetism,
  title={Ferromagnetism and topology of the higher flat band in a fractional Chern insulator},
  author={Park, Heonjoon and Cai, Jiaqi and Anderson, Eric and Zhang, Xiao-Wei and Liu, Xiaoyu and Holtzmann, William and Li, Weijie and Wang, Chong and Hu, Chaowei and Zhao, Yuzhou and others},
  journal={Nat. Phys.},
  pages={1--7},
  year={2025},
  publisher={Nature Publishing Group UK London}
}

@article{li2025universal,
  title={Universal Magnetic Phases in Twisted Bilayer {MoTe$_2$}},
  author={Li, Weijie and Redekop, Evgeny and Beach, Christiano Wang and Zhang, Canxun and Zhang, Xiaowei and Liu, Xiaoyu and Holtzmann, Will and Hu, Chaowei and Anderson, Eric and Park, Heonjoon and others},
  journal={arXiv preprint arXiv:2507.22354},
  year={2025}
}

@article{chen2020tunable,
  title={Tunable correlated Chern insulator and ferromagnetism in a moir{\'e} superlattice},
  author={Chen, Guorui and Sharpe, Aaron L and Fox, Eli J and Zhang, Ya-Hui and Wang, Shaoxin and Jiang, Lili and Lyu, Bosai and Li, Hongyuan and Watanabe, Kenji and Taniguchi, Takashi and others},
  journal={Nature},
  volume={579},
  number={7797},
  pages={56--61},
  year={2020},
  publisher={Nature Publishing Group UK London}
}

@article{tschirhart2021imaging,
  title={Imaging orbital ferromagnetism in a moir{\'e} Chern insulator},
  author={Tschirhart, CL and Serlin, Marec and Polshyn, Hryhoriy and Shragai, Avi and Xia, Zhengchao and Zhu, Jiacheng and Zhang, Yuxuan and Watanabe, Kenji and Taniguchi, Takashi and Huber, ME and others},
  journal={Science},
  volume={372},
  number={6548},
  pages={1323--1327},
  year={2021},
  publisher={American Association for the Advancement of Science}
}

@article{zeng2023thermodynamic,
  title={Thermodynamic evidence of fractional Chern insulator in moir{\'e} MoTe2},
  author={Zeng, Yihang and Xia, Zhengchao and Kang, Kaifei and Zhu, Jiacheng and Kn{\"u}ppel, Patrick and Vaswani, Chirag and Watanabe, Kenji and Taniguchi, Takashi and Mak, Kin Fai and Shan, Jie},
  journal={Nature},
  volume={622},
  number={7981},
  pages={69--73},
  year={2023},
  publisher={Nature Publishing Group UK London}
}

@article{lu2025extended,
  title={Extended quantum anomalous Hall states in graphene/hBN moir{\'e} superlattices},
  author={Lu, Zhengguang and Han, Tonghang and Yao, Yuxuan and Hadjri, Zach and Yang, Jixiang and Seo, Junseok and Shi, Lihan and Ye, Shenyong and Watanabe, Kenji and Taniguchi, Takashi and others},
  journal={Nature},
  volume={637},
  number={8048},
  pages={1090--1095},
  year={2025},
  publisher={Nature Publishing Group UK London}
}

@article{wang2024phase,
  title = {Phase diagram of twisted bilayer ${\mathrm{MoTe}}_{2}$ in a magnetic field with an account for the electron-electron interaction},
  author = {Wang, Minxuan and Wang, Xiaoyu and Vafek, Oskar},
  journal = {Phys. Rev. B},
  volume = {110},
  issue = {20},
  pages = {L201107},
  numpages = {6},
  year = {2024},
  month = {Nov},
  publisher = {American Physical Society},
  doi = {10.1103/PhysRevB.110.L201107},
  url = {https://link.aps.org/doi/10.1103/PhysRevB.110.L201107}
}

@misc{kang2025orbital,
      title={Orbital magnetization and magnetic susceptibility of interacting electrons}, 
      author={Jian Kang and Minxuan Wang and Oskar Vafek},
      year={2025},
      eprint={2509.20626},
      archivePrefix={arXiv},
      url={https://arxiv.org/abs/2509.20626}, 
}

@article{qiu2023interaction,
  title={Interaction-driven topological phase diagram of twisted bilayer {MoTe$_2$}},
  author={Qiu, Wen-Xuan and Li, Bohao and Luo, Xun-Jiang and Wu, Fengcheng},
  journal={Phys. Rev. X},
  volume={13},
  number={4},
  pages={041026},
  year={2023},
  publisher={APS}
}

@article{fan2024intrinsic,
  title={Intrinsic dipole Hall effect in twisted MoTe2: magnetoelectricity and contact-free signatures of topological transitions},
  author={Fan, Feng-Ren and Xiao, Cong and Yao, Wang},
  journal={Nat. Commun.},
  volume={15},
  number={1},
  pages={7997},
  year={2024},
  publisher={Nature Publishing Group UK London}
}

@article{zhu2024layer,
  title={Layer Hall counterflow as a model probe of magic-angle twisted bilayer graphene},
  author={Zhu, Jihang and Zhai, Dawei and Xiao, Cong and Yao, Wang},
  journal={Phys. Rev. B},
  volume={109},
  number={15},
  pages={155114},
  year={2024},
  publisher={APS}
}

@article{guinea2018electrostatic,
  title={Electrostatic effects, band distortions, and superconductivity in twisted graphene bilayers},
  author={Guinea, Francisco and Walet, Niels R},
  journal={Proc. Natl. Acad. Sci. U.S.A},
  volume={115},
  number={52},
  pages={13174--13179},
  year={2018},
  publisher={National Academy of Sciences}
}

@article{zhu2024weak,
  title={Weak-coupling theory of magic-angle twisted bilayer graphene},
  author={Zhu, Jihang and Torre, Iacopo and Polini, Marco and MacDonald, Allan H},
  journal={Phys. Rev. B},
  volume={110},
  number={12},
  pages={L121117},
  year={2024},
  publisher={APS}
}




\clearpage
\onecolumngrid
\setcounter{equation}{0}
\setcounter{figure}{0}
\setcounter{table}{0}
\setcounter{page}{1}
\setcounter{section}{0}
\setcounter{secnumdepth}{3}
\makeatletter
\renewcommand{\theequation}{S\arabic{equation}}
\renewcommand{\thefigure}{S\arabic{figure}}
\renewcommand{\thesection}{S\arabic{section}}
\makeatother
\begin{center}
\textbf{\large Supplemental Material for ``Orbital Magnetization of Correlated Phases in Twisted Bilayer Transition Metal Dichalcogenides''}

\bigskip

Xiaoyu Liu,$^1$ Chong Wang,$^1$ Haoran Chen,$^1$ Xiao-Wei Zhang,$^1$ Ting Cao,$^1$ and Di Xiao$^{1,2,3}$

\smallskip

\textit{$^1$Department of Materials Science and Engineering, University of Washington, Seattle, WA 98195, USA}

\textit{$^2$Department of Physics, University of Washington, Seattle, WA 98195, USA}

\textit{$^3$Pacific Northwest National Laboratory, Richland, WA, USA}

\end{center}

\section{Detailed Derivation of Orbital Magnetization in Hartree-Fock Approximation}
Orbital magnetization at zero temperature is defined as
\begin{equation}
 \bm{M} = -\frac{1}{V}\left(\frac{\partial K}{\partial B}\right)_\mu,
\end{equation}
where $K = E - \mu N$ is the grand potential at zero temperature, $\mu$ is the chemical potential, $N$ is the particle number, $V$ is the volume, and $E$ is the energy of the system. For simplicity, we assume the system is an insulator. The derivation for metallic systems is similar, although the Hartree-Fock approximation is known to be relatively less accurate for metallic phases. {The Hamiltonian of an interacting system can be written as $\hat{H} = \hat{H}_0 + \hat{H}_{\rm int}$, where $\hat{H}_0$ denotes the non-interacting Hamiltonian and $\hat{H}_\text{int}$ accounts for the Coulomb interaction. 
Within the Hartree-Fock approximation, the four-fermion operator $\hat{H}_\text{int}$ is reduced to an effective two-fermion operator $\hat{H}_\text{MF}[\rho]$, expressed as a functional of the one-body reduced density matrix $\rho$. 
We denote $\hat{H}_\text{HF}=\hat{H}_0+\hat{H}_\text{MF}$ as the Hartree-Fock Hamiltonian. 
Solving the Hartree-Fock problem requires a self-consistency loop, and the quasi-orbitals obtained from this procedure are referred to as Hartree–Fock orbitals.}

A uniform magnetic field is difficult to deal with, in that it is a singular perturbation. The singular nature can be observed from the fact that the vector potential $\bm{A} (\bm{r})$ is unbounded in the space if the magnetic field is uniform. To circumvent this problem, we introduce a periodic
magnetic field:
\begin{eqnarray}
  \bm{B} (\bm{r}) & = & B \cos (\bm{q} \cdot \bm{r})
  \bm{e}_z,
\end{eqnarray}
where $\bm{e}_z$ is the unit vector in the $+ z$ direction and
$\bm{q}= q \bm{e}_y$. In the Landau gauge,
\begin{eqnarray}\label{A_supp}
  \bm{A} (\bm{r}) & = & - B \frac{\sin (\bm{q} \cdot
  \bm{r})}{q} \bm{e}_x.
\end{eqnarray}
The perturbation to the Hamiltonian is
\begin{eqnarray}\label{V_B_supp}
  \hat{V}_B & = & \frac{e}{2} [\hat{\bm{v}}_0 \cdot \bm{A}
  (\hat{\bm{r}}) +\bm{A} (\hat{\bm{r}}) \cdot
  \hat{\bm{v}}_0].
\end{eqnarray}
Assuming there is nothing infinitely nonlocal, in the long-wave limit ($q \to 0$), the variation in the local density of $K$ is related to orbital magnetization by~\cite{Shi_Quantum_2007}
\begin{equation}\label{eq:orbmom_density}
\delta K(\bm{r})=-\bm{M}\cdot \bm{B}(\bm{r}). 
\end{equation}
In this way, the orbital magnetization can be calculated through
\begin{equation}\label{eq:orbmom_integration}
    M_z=-\frac{2}{VB}\lim_{q \to 0}\int d\bm{r}\delta K(\bm{r})\textrm{cos}(\bm{q}\cdot\bm{r}).
\end{equation}

In Hartree-Fock approximation, the mean-field Hartree-Fock Hamiltonian (without magnetic field) $\hat{H}_{\mathrm{HF}}$ has two parts: the non-interacting Hamiltonian $\hat{H}_0$, and the mean-field decomposition of the electron-electron interactions $\hat{H}_{\rm MF}$ which inlcludes both the Hartree Hamiltonian and the Fock Hamiltonian. The Hartree-Fock orbitals (without magnetic field) are denoted as $|\phi_{n\bm{k}}\rangle$. When the periodic magnetic field is turned on, $|\phi_{n\bm{k}}\rangle$ changes by $|\delta\phi_{n\bm{k}}\rangle$. Correspondingly, the energy of the system changes by ($f_{n\bm{k}}$ is the occupation number)
\begin{equation}
    \delta E = \sum_n f_{n\bm{k}} \left[\langle \phi_{n\bm{k}}| \hat{V}_B| \phi_{n\bm{k}}\rangle + \left(\langle \phi_{n\bm{k}} | \hat{H}_{\mathrm{HF}}| \delta \phi_{n\bm{k}}\rangle + c.c.\right) \right].
\end{equation}
Therefore, it makes sense to define the local density of $K$ as
\begin{equation}
    \begin{aligned}
        \delta K(\bm{r}) & = \delta K_1 (\bm{r}) + \delta K_2 (\bm{r}) \\
        \delta K_1 (\bm{r}) & =
        \frac{1}{2}\left[\sum_{n\bm{k}} f_{n\bm{k}}\langle\phi_{n\bm{k}} | (\hat{H}_{\rm HF} - \mu\hat{N})\delta(\hat{\bm{r}}-\bm{r})  
         + \delta(\hat{\bm{r}}-\bm{r}) (\hat{H}_{\rm HF} - \mu\hat{N})| \delta \phi_{n\bm{k}}\rangle + c.c.\right] \\
        \delta K_2 (\bm{r}) & = \frac{1}{2} \sum_{n\bm{k}} f_{n\bm{k}} \langle \phi_{n\bm{k}}| \hat{V}_B \delta(\hat{\bm{r}}-\bm{r}) + \delta(\hat{\bm{r}}-\bm{r}) \hat{V}_B|\phi_{n\bm{k}}\rangle.
    \end{aligned} \label{eq:local_energy_supp}
\end{equation}
{
$\delta K_2 (\bm{r})$ in Eq.~(\ref{eq:local_energy_supp}) does not contribute to $M_z$ after the integration in Eq.~\eqref{eq:orbmom_integration}. More specifically, using Eq.~\eqref{V_B_supp} and Eq.~\eqref{A_supp}, the contribution of the $\delta K_2 (\bm{r})$ can be decomposed into four terms. As an illustration, consider the first term:
\begin{equation}
\begin{aligned}
    & \frac{e}{2 q V} \lim_{q \to 0} \int d \bm{r} \sum_{n\bm{k}} f_{n\bm{k}} \langle \phi_{n\bm{k}}| \hat{v}_{0,x} \sin (\bm{q} \cdot \hat{\bm{r}}) \delta(\hat{\bm{r}}-\bm{r})|\phi_{n\bm{k}}\rangle \cos (\bm{q} \cdot \bm{r}) \\
    = & \frac{e}{2 q V} \lim_{q \to 0} \sum_{n\bm{k}} f_{n\bm{k}} \langle \phi_{n\bm{k}}| \hat{v}_{0,x} \sin (\bm{q} \cdot \hat{\bm{r}}) \cos (\bm{q} \cdot \bm{\hat{r}}) |\phi_{n\bm{k}}\rangle \\
    = & \frac{e}{4 q V} \lim_{q \to 0} \sum_{n\bm{k}} f_{n\bm{k}} \langle \phi_{n\bm{k}}| \hat{v}_{0,x} \sin (2 \bm{q} \cdot \hat{\bm{r}}) |\phi_{n\bm{k}}\rangle.
\end{aligned}
\end{equation}
The last expression vanishes since the momentum introduced by $\sin (2 \bm{q} \cdot \hat{\bm{r}})$ cannot be compensated. The other three terms can be treated in the same way and also vanish. This indicates that $\delta K_2 (\bm{r})$ does not have a component of the form $\cos (\bm{q} \cdot \bm{r})$.

In addition, $\delta K_2 (\bm{r})$ does not have a component of the form $\sin (\bm{q} \cdot \bm{r})$ either. This can be expected from Eq.~\eqref{eq:orbmom_density}, which indicates that a magnetic field of the form $\cos (\bm{q} \cdot \bm{r})$ can only induce $\delta K (\bm{r})$ of the form $\cos (\bm{q} \cdot \bm{r})$. Explicitly,
\begin{equation}
\begin{aligned}
    & \frac{e}{2 q V} \lim_{q \to 0} \int d \bm{r} \sum_{n\bm{k}} f_{n\bm{k}} \langle \phi_{n\bm{k}}| \hat{v}_{0,x} \sin (\bm{q} \cdot \hat{\bm{r}}) \delta(\hat{\bm{r}}-\bm{r})|\phi_{n\bm{k}}\rangle \sin (\bm{q} \cdot \bm{r}) \\
    = & \frac{e}{2 q V} \lim_{q \to 0} \sum_{n\bm{k}} f_{n\bm{k}} \langle \phi_{n\bm{k}}| \hat{v}_{0,x} \sin (\bm{q} \cdot \hat{\bm{r}}) \sin (\bm{q} \cdot \bm{\hat{r}}) |\phi_{n\bm{k}}\rangle \\
    = & \frac{e}{4 q V} \lim_{q \to 0} \sum_{n\bm{k}} f_{n\bm{k}} \langle \phi_{n\bm{k}}| \hat{v}_{0,x} |\phi_{n\bm{k}}\rangle - \frac{e}{4 q V} \lim_{q \to 0} \sum_{n\bm{k}} f_{n\bm{k}} \langle \phi_{n\bm{k}}| \hat{v}_{0,x} \cos (2 \bm{q} \cdot \hat{\bm{r}}) |\phi_{n\bm{k}}\rangle.
\end{aligned}
\end{equation}
In the last line of the above equation, the first term vanishes since no net current can be carried by an equilibrium state, while the second term vanishes again due to momentum mismatch.
}

Therefore, we focus on $\delta K_1 (\bm{r})$. The variation of the Hartree-Fock orbitals contains two parts:
\begin{equation}
|\delta\phi_{n\bm{k}}^{(1)}\rangle=\sum_{n',\bm{k}'}'\frac{|\phi_{n'\bm{k}'}\rangle \langle \phi_{n'\bm{k}'}| \hat{V}_{\rm B} | \phi_{n\bm{k}} \rangle}{\epsilon_{n\bm{k}} - \epsilon_{n'\bm{k}'}},
\end{equation}
and
\begin{equation}
|\delta\phi_{n\bm{k}}^{(2)}\rangle=\sum_{n',\bm{k}'}'\frac{|\phi_{n'\bm{k}'}\rangle \langle \phi_{n'\bm{k}'}| \delta \hat{H}_{\rm HF} | \phi_{n\bm{k}} \rangle}{\epsilon_{n\bm{k}} - \epsilon_{n'\bm{k}'}},
\end{equation}
where the summation excludes the term with $n = n'$ and $\bm{k} = \bm{k}'$, $\epsilon_{n\bm{k}}$ is the quasi-energy of the Hartree-Fock orbital, and $\delta \hat{H}_{\rm HF}$ is the self-consistent change of the Hartree-Fock Hamiltonian. $|\delta\phi_{n\bm{k}}^{(1)}\rangle$ corresponds to
\begin{equation}
\begin{aligned}
    \delta K_1^{(1)} (\bm{r})
    & = \textrm{Re}\left[\sum_{n \bm{k}}\sum_{n' \bm{k}'}^{'} f_{n \bm{k}} \big\langle \phi_{n\bm{k}} \big| 
    (\hat{H}_{\rm HF}-\mu\hat{N}) \delta(\hat{\bm{r}}-\bm{r}) + \delta(\hat{\bm{r}}-\bm{r})(\hat{H}_{\rm HF}-\mu\hat{N})
        \big| \phi_{n'\bm{k}'} \big\rangle \frac{\langle \phi_{n'\bm{k}'} | \hat{V}_B | \phi_{n\bm{k}} \rangle}{\epsilon_{n\bm{k}} - \epsilon_{n'\bm{k}'}} \right], \\
    &=-\frac{eB}{4q} \textrm{Im}\sum_{n\bm{k}}\sum_{n'\bm{k}'}^{'}f_{n\bm{k}}(\epsilon_{n\bm{k}}+\epsilon_{n'\bm{k}'}-2\mu)\langle \phi_{n\bm{k}}| \delta(\hat{\bm{r}}-\bm{r})|\phi_{n'\bm{k}'} \rangle 
    \left[\frac{\langle \phi_{n'\bm{k}'} | \hat{v}_{0,x} e^{i\bm{q}\cdot \hat{\bm{r}}} + e^{i\bm{q}\cdot \hat{\bm{r}}} \hat{v}_{0,x} | \phi_{n\bm{k}}\rangle}{\epsilon_{n\bm{k}} - \epsilon_{n'\bm{k}'}} - (\bm{q}\rightarrow -\bm{q})\right].\\
\end{aligned}
\end{equation}
For the matrix element $\langle \phi_{n'\bm{k}'} | \hat{v}_{0,x} e^{i\bm{q}\cdot \hat{\bm{r}}} + e^{i\bm{q}\cdot \hat{\bm{r}}} \hat{v}_{0,x}| \phi_{n\bm{k}}\rangle$ to be nonzero, $\bm{k}'$ needs to be set to $\bm{k} + \bm{q}$ due to momentum conservation, leading to
\begin{equation}
\begin{aligned}
    \delta K_1^{(1)} (\bm{r})
    &=-\frac{eB}{4q} \textrm{Im}\sum_{nn'\bm{k}} f_{n\bm{k}}\left[(\epsilon_{n\bm{k}}+\epsilon_{n'\bm{k}+\bm{q}}-2\mu)\langle \phi_{n\bm{k}}| \delta(\hat{\bm{r}}-\bm{r})|\phi_{n'\bm{k}+\bm{q}} \rangle 
    \frac{\langle \phi_{n'\bm{k}+\bm{q}} | \hat{v}_{0,x} e^{i\bm{q}\cdot \hat{\bm{r}}} + e^{i\bm{q}\cdot \hat{\bm{r}}} \hat{v}_{0,x}| \phi_{n\bm{k}}\rangle}{\epsilon_{n\bm{k}} - \epsilon_{n'\bm{k}+\bm{q}}} - (\bm{q}\rightarrow -\bm{q})\right],\\
    &= -\frac{eB}{4q}\textrm{Im}\sum_{nn'\bm{k}} 
    (\epsilon_{n\bm{k}}+\epsilon_{n'\bm{k}+\bm{q}}-2\mu)
    \langle \phi_{n\bm{k}}| \delta(\hat{\bm{r}}-\bm{r})|\phi_{n'\bm{k}+\bm{q}} \rangle 
    \langle \phi_{n'\bm{k}+\bm{q}}| \hat{v}_{0,x} e^{i\bm{q}\cdot \hat{\bm{r}}} +  e^{i\bm{q}\cdot \hat{\bm{r}}} \hat{v}_{0,x}| \phi_{n\bm{k}}\rangle
    \frac{f_{n\bm{k}}-f_{n'\bm{k}+\bm{q}}}{\epsilon_{n\bm{k}}-\epsilon_{n'\bm{k}+\bm{q}}}.
\end{aligned}
\end{equation}
In the derivation, we have shifted $\bm{k}-\bm{q} \to \bm{k}$ and $\bm{k} \to \bm{k}+\bm{q}$ for the $(\bm{q} \to -\bm{q})$ part in the bracket. This substitution is legitimate, since the summation over $\bm{k}$ is over the whole Brillouin zone.

The orbital magnetization contributed by $\delta K_1^{(1)} (\bm{r})$ is thus
\begin{equation}
\begin{aligned}
    M_z^{(1)} &= \lim_{q \to 0} \frac{e}{2qV} \textrm{Im} \sum_{nn'\bm{k}} (\epsilon_{n\bm{k}}+\epsilon_{n'\bm{k}+\bm{q}}-2\mu)
    \langle \phi_{n\bm{k}}| \cos(\bm{q}\cdot\hat{\bm{r}})|\phi_{n'\bm{k}+\bm{q}} \rangle 
    \langle \phi_{n'\bm{k}+\bm{q}}| \hat{v}_{0,x} e^{i\bm{q}\cdot \hat{\bm{r}}} +  e^{i\bm{q}\cdot \hat{\bm{r}}} \hat{v}_{0,x}| \phi_{n\bm{k}}\rangle
    \frac{f_{n\bm{k}}-f_{n'\bm{k}+\bm{q}}}{\epsilon_{n\bm{k}}-\epsilon_{n'\bm{k}+\bm{q}}}, \\
    &=\lim_{q \to 0} \frac{e}{4qV} \textrm{Im} \sum_{nn'\bm{k}} (\epsilon_{n\bm{k}}+\epsilon_{n'\bm{k}+\bm{q}}-2\mu)
    \langle u_{n\bm{k}}|u_{n'\bm{k}+\bm{q}} \rangle 
    \langle u_{n'\bm{k}+\bm{q}}| \hat{v}_{0,x}(\bm{k}+\bm{q})  +   \hat{v}_{0,x}(\bm{k})| u_{n\bm{k}}\rangle
    \frac{f_{n\bm{k}}-f_{n'\bm{k}+\bm{q}}}{\epsilon_{n\bm{k}}-\epsilon_{n'\bm{k}+\bm{q}}},
\end{aligned}
\label{eq:Mz1_supp}
\end{equation}
where $\hat{v}_{0,x}(\bm{k}) = e^{-i \bm{k} \cdot \bm{r}} \hat{v}_{0,x} e^{i \bm{k} \cdot \bm{r}} $, and $ u_{n\bm{k}}$ is the cell-periodic part of $\phi_{n\bm{k}}$.
Notice that the above equation has the same structure compared with Eq.~(11) in~\cite{Shi_Quantum_2007}. The only difference is that here the orbitals $u_{n\bm{k}}$ are Hartree-Fock orbitals, and the velocity operator $\hat{v}_{0,x}(\bm{k})$ is the velocity operator corresponding to $\hat{H}_0$.

We now discuss $M_z^{(2)}$, which is contributed by $|\delta\phi_{n\bm{k}}^{(2)}\rangle$ from the self-consistent change of the Hartree-Fock Hamiltonian $\delta \hat{H}_{\rm HF}$. $\delta \hat{H}_{\rm HF}$ is fully contributed by the change of $\hat{H}_{\rm MF}$. Generally, for an external perturbation, $\delta \hat{H}_{\rm HF}$ needs to be obtained in a self-consistent fashion. However, since eventually we will take the limit $q \to 0$, we only need to analyze the behavior of $\delta \hat{H}_{\rm HF}$ in the small $q$ limit. In the small $q$ limit, $\bm{A}$ varies slowly in real space. On the other hand, when $\bm{A}$ is simply a gauge transformation, and the mean-field Hamiltonian simply transforms as
\begin{equation}
    \hat{H}_{\rm MF}(\bm{A}) = e^{-i e \bm{A} \cdot \hat{\bm{r}}} \hat{H}_{\rm MF}(\bm{A}=\bm{0}) e^{i e \bm{A} \cdot \hat{\bm{r}}},
\end{equation}
such that $\delta \hat{H}_{\rm HF} = e\hat{\bm{v}}_{\rm MF} \cdot \bm{A}$, where $\hat{\bm{v}}_{\rm MF} = -i [\hat{\bm{r}}, \hat{H}_{\rm MF}] / \hbar$. Assuming locality in Hartree-Fock calculations, for a slowly varying vector potential $\bm{A}(\bm{r})$, $\delta \hat{H}_{\rm HF} \approx e[\hat{\bm{v}}_{\rm MF} \cdot \bm{A}(\hat{\bm{r}}) + \bm{A}(\hat{\bm{r}}) \cdot \hat{\bm{v}}_{\rm MF}] / 2 $. Here, we have symmetrized $\hat{\bm{v}}_{\rm MF}$ and $\bm{A}(\hat{\bm{r}})$ to ensure that $\delta \hat{H}_{\rm HF}$ is Hermitian. Other symmetrization schemes are possible, but will not affect the final result. With the knowledge of $\delta \hat{H}_{\rm HF}$, it is straightforward to show that $M_z^{(2)}$ has the same form as $M_z^{(1)}$ in Eq.~(\ref{eq:Mz1_supp}), but with $\hat{\bm{v}}_0$ replaced by $\hat{\bm{v}}_{\rm MF}$.

Putting everything together,
\begin{equation}
    M_z
    =\lim_{q \to 0} \frac{e}{4qV} \textrm{Im} \sum_{nn'\bm{k}} (\epsilon_{n\bm{k}}+\epsilon_{n'\bm{k}+\bm{q}}-2\mu)
    \langle u_{n\bm{k}}|u_{n'\bm{k}+\bm{q}} \rangle 
    \langle u_{n'\bm{k}+\bm{q}}| \hat{v}_{\mathrm{HF},x}(\bm{k}+\bm{q})  +   \hat{v}_{\mathrm{HF},x}(\bm{k})| u_{n\bm{k}}\rangle
    \frac{f_{n\bm{k}}-f_{n'\bm{k}+\bm{q}}}{\epsilon_{n\bm{k}}-\epsilon_{n'\bm{k}+\bm{q}}},
\label{eq:Mz}
\end{equation}
where $\hat{\bm{v}}_{\rm HF} = \hat{\bm{v}}_0 + \hat{\bm{v}}_{\rm MF} = -i [\hat{\bm{r}}, \hat{H}_{\rm HF}] / \hbar$. 
{
In the above expression, the $n = n'$ terms vanish due to $f_{n \bm{k}} = f_{n \bm{k} + \bm{q}}$, as an insulator has been assumed. For $n \neq n'$, in the limit of $q \to 0$, the $1/q$ factor has to be paired with $\langle u_{n\bm{k}}|u_{n'\bm{k}+\bm{q}} \rangle$ in the L'H\^{o}pital's rule, leading to
\begin{equation}
\begin{aligned}
    M_z
    =& \frac{e}{2V} \textrm{Im} \sum_{n\neq n'\bm{k}} (\epsilon_{n\bm{k}}+\epsilon_{n'\bm{k}}-2\mu)
    \langle u_{n\bm{k}}|\partial_{k_y} u_{n'\bm{k}} \rangle 
    \langle u_{n'\bm{k}}| \hat{v}_{\mathrm{HF},x}(\bm{k}) | u_{n\bm{k}}\rangle
    \frac{f_{n\bm{k}}-f_{n'\bm{k}}}{\epsilon_{n\bm{k}}-\epsilon_{n'\bm{k}}}\\
    = & \frac{e}{2\hbar V} \textrm{Im} \sum_{n\neq n' \bm{k}} 
    \bigg[(f_{n\bm{k}} - f_{n'\bm{k}}) (\epsilon_{n\bm{k}}+\epsilon_{n'\bm{k}}-2\mu)
     \langle  u_{n\bm{k}}| \partial_{k_y} u_{n'\bm{k}}\rangle  \langle u_{n'\bm{k}} | \partial_{k_x} u_{n\bm{k}}\rangle 
      ]\\
    = & \frac{e}{2\hbar V}  \textrm{Im} \sum_{n\bm{k}} f_{n\bm{k}} \bigg[
    \langle\partial_{k_x} u_{n\bm{k}}|
    (\hat{H}_{\textrm{HF}}+ \epsilon_{n\bm{k}} - 2\mu) | \partial_{k_y} u_{n\bm{k}}\rangle - 
    \langle\partial_{k_y} u_{n\bm{k}}|
    (\hat{H}_{\textrm{HF}}+ \epsilon_{n\bm{k}} - 2\mu) | \partial_{k_x} u_{n\bm{k}}\rangle \bigg] \\
    = & \frac{e}{2\hbar V} \sum_{n\bm{k}} f_{n\bm{k}} \bigg[
    \langle\partial_{k_x} u_{n\bm{k}}|
    (\hat{H}_{\textrm{HF}}+ \epsilon_{n\bm{k}} - 2\mu) | \partial_{k_y} u_{n\bm{k}}\rangle - 
    \langle\partial_{k_y} u_{n\bm{k}}|
    (\hat{H}_{\textrm{HF}}+ \epsilon_{n\bm{k}} - 2\mu) | \partial_{k_x} u_{n\bm{k}}\rangle \bigg] \\
    = & -\frac{i e}{2\hbar V} \sum_{n\bm{k}}f_{n\bm{k}}
\bigg\langle\frac{\partial u_{n\bm{k}}}{\partial\bm{k}}\bigg|
\times[\hat{H}_{\rm HF}(\bm{k})+\epsilon_{n\bm{k}}-2\mu]
\bigg|\frac{\partial u_{n\bm{k}}}{\partial\bm{k}}\bigg\rangle,
\end{aligned}
\end{equation}
where $\hat{H}_{\rm HF}(\bm{k}) = e^{-i \bm{k} \cdot \bm{r}} \hat{H}_{\rm HF} e^{i \bm{k} \cdot \bm{r}} $. In other words, orbital magnetization in Hartree-Fock approximation can be evaluated with Hartree-Fock Hamiltonian and orbitals as if the system has no interactions. In the derivation, we have used the relation between the velocity operator and the Berry connection $\langle u_{n'\bm{k}}| \hat{\bm{v}}_{\mathrm{HF}}(\bm{k}) | u_{n\bm{k}}\rangle =  (\epsilon_{n\bm{k}}-\epsilon_{n'\bm{k}})\langle u_{n'\bm{k}} | \partial_{\bm{k}} u_{n\bm{k}}\rangle / \hbar$.}

We note that $\delta \hat{H}_{\rm HF} \approx e[\hat{\bm{v}}_{\rm MF} \cdot \bm{A}(\hat{\bm{r}}) + \bm{A}(\hat{\bm{r}}) \cdot \hat{\bm{v}}_{\rm MF}] / 2 $ can be viewed as an expansion of $\delta \hat{H}_{\rm HF}$ in $q$ around $q = 0$. $\bm{A}(\hat{\bm{r}})$ is proportional to $1 / q$. The next order contribution to $\delta \hat{H}_{\rm HF}$ is expected to be proportional to gradients of $\bm{A}$ and is ${\rm O}(1)$. However, in taking the limit of $q \to 0$, a $1/q$ has to be paired with $\langle u_{n\bm{k}}|u_{n'\bm{k}+\bm{q}} \rangle$ in Eq.~(\ref{eq:Mz}) in the L'H\^{o}pital's rule. Therefore, an ${\rm O}(1)$ contribution to $\delta \hat{H}_{\rm HF}$ does not contribute to the expression of orbital magnetization. {In addition, the form of $\delta \hat{H}_{\rm HF}$ only relies on gauge invariance, such that the above analysis is applicable to self-consistent Hartree calculations, which is sometimes employed to model graphene-based moir\'e superlattices~\cite{guinea2018electrostatic,zhu2024weak}.}

{The simple extension of existing formulas for orbital magnetization is partly due to the fact that it is a first-order derivative of a thermodynamic quantity. For response functions involving higher-order derivatives, or those of a nonequilibrium nature~\cite{fan2024intrinsic,zhu2024layer}, the corresponding Hartree–Fock expressions remain to be explored.}

{
\section{Field theory description}
Under Hartree-Fock approximation, the mean-field Hamiltonian can be written as $\hat{H}_{\mathrm{HF}}=i\hbar\omega_n-\hat{G}^{-1}$, where $\hat{G}$ is the Green's function and $i\omega_n$ is the Fermionic Matsubara frequency. Under a electromagnetic field, up to the linear order, it couples to the vector potential in the form of $\hat{H}_{\mathrm{HF}}^{A}=\hat{H}_{\mathrm{HF}}^0+A_\mu \hat{\Gamma}^{\mu}$. To this end, we are to determine $\hat{\Gamma}^{\mu}(1,2;3)\equiv-\left.\delta \hat{G}^{-1}(1,2) / \delta A_\mu(3)\right|_{A=0}$. The function is called a vertex function. Here $1$ denotes $x_{1}\equiv(\bm{r}_{1},t_{1})$ being 4-dimensional coordinates and $\mu$ denotes their components. As will be shown later, the vertex function is closely related to the mean-field velocity operator.

On the other hand, charge conservation condition requires that, under an $U(1)$ gauge transformation ${A}\rightarrow{A}+{\partial} \alpha$, where $\partial\equiv(\partial_t,\bm{\nabla})$, the Green's function should transform covariantly, i.e., 
\begin{equation}
\begin{aligned}
\hat{G}^{-1}[A+{\partial} \alpha](1,2)=e^{-e\alpha(1) / i\hbar}\,\hat{G}^{-1}[A]\,e^{e \alpha (2) / i\hbar}.
\end{aligned}\label{eq:gauge_invariance}
\end{equation}
For an infinitesimal $\alpha$, we can expand both sides in terms of $\alpha$. 
Up to first order of $\alpha$, the left-hand side of \eqref{eq:gauge_invariance} equals
\begin{equation}
\begin{aligned}
	\hat{G}^{-1}[A+{\partial} \alpha](1,2) &=
    \hat{G}^{-1}[A](1,2) + \frac{\delta \hat{G}^{-1}(1,2)}{\delta A(3)} {\partial}\alpha(3)\\
	&= \hat{G}^{-1}[A](1,2) - {\partial}_3\left(\frac{\delta \hat{G}^{-1}(1,2)}{\delta A(3)}\right) \alpha(3),
\end{aligned}
\end{equation}
where from the first to the second line, we performed integration by parts over $3$. 
The right-hand side of Eq.~\eqref{eq:gauge_invariance} equals
\begin{equation}
\begin{aligned}
	e^{-\frac{e}{i\hbar}\alpha(1)}\,\hat{G}^{-1}[A]\,e^{\frac{e}{i\hbar}\alpha(2)}
	&=\hat{G}^{-1}[A](1,2) + \frac{e}{i\hbar} \big(\alpha(2)-\alpha(1)\big) \hat{G}^{-1}[A](1,2)\\
	&=\hat{G}^{-1}[A](1,2)+\frac{e}{i\hbar} \hat{G}^{-1}[A](1,2)\big(\delta(2-3)-\delta(1-3)\big)\alpha(3).
\end{aligned}
\end{equation}
By comparing the two equations and further Fourier transforming to the 4-dimensional momentum space, we get the Ward-Takahashi identity
\begin{equation}
	q_\mu\hat{\Gamma}^\mu(k+q, k)=\frac{e}{\hbar}\left(\hat{G}^{-1}(k+q)-\hat{G}^{-1}(k)\right).
\end{equation}

If the self energy $\hat{\Sigma}(i\omega_n)=i\hbar\omega_n-\hat{G}^{-1}(i\omega_n)-\hat{H}_0$ of a system is frequency-independent, in the long-wavelength limit, the spatial part of the Ward-Takahashi identity gives
\begin{equation}
	\hat{\bm{\Gamma}}(\bm{k})=-\frac{e}{\hbar}\partial_{\bm{k}}\left(\hat{H}_{0}(\bm{k})+\hat{\Sigma}(\bm{k})\right).
\end{equation}
For a non-interacting system, this gives $\hat{\bm{\Gamma}}= -\frac{e}{m_e} \hat{p}_0$ being the opposite bare current operator, and we obtain the minimal coupling Hamiltonian $\hat{H}_0^A=\hat{H}_0 - \frac{e}{m_e} \hat{p}_0 A + O(A^2)$.
For a system with interaction or external perturbation such as disorder, these effects will lead to vertex corrections. 
Under Hartree-Fock approximation, the corrected vertex becomes
\begin{equation}
	\hat{\bm{\Gamma}}_{\mathrm{HF}}=-\frac{e}{\hbar}\partial_{\bm{k}}\hat{H}_{\textrm{HF}}(\bm{k})=-e\hat{\bm{v}}_{\textrm{HF}}(\bm{k}).
\end{equation}
If we further divide the Hartree Fock Hamiltonian into bare, Hartree and Fock terms as $\hat{H}_{\textrm{HF}} = \hat{H}_0+\hat{\Sigma}_H+\hat{\Sigma}_F$, we will find the vertex correction originates solely from the Fock term. This is because the Hartree potential, being a local potential, is $\bm{k}$-independent. As a result, we have $\hat{\bm{v}}_{\mathrm{HF}}=\hat{\bm{v}}_0+\partial_{\bm{k}}\hat{\Sigma}_F(\bm{k})/\hbar$. 
}




\section{Kane-Mele-Hubbard model under magnetic field}
\begin{figure}[h]
\centering
\includegraphics[width=0.5\columnwidth]{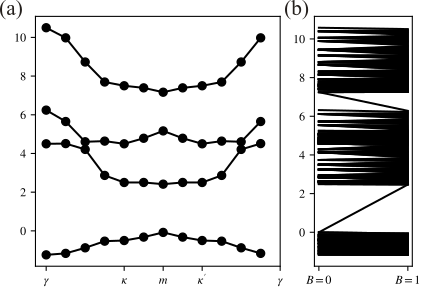}
\caption{(a) Hartree-Fock spectrum without magnetic field. $k$-mesh is $12\times12$. (b) Change of energy levels of a $12x12$ supercell without and with magnetic field. $t_1=1, t_2=1/3, \theta=\pi/3, U=10, V=1$. \label{fig:levels}}
\end{figure}

A typical Hartree–Fock band structure of the extended KMH model, Eq.~(\ref{eq:kmh}), is shown in Fig.~\ref{fig:levels}(a).
At filling $\nu=-1$, the system is spontaneously valley (spin) polarized.
The onsite Coulomb interaction $U$ splits the two valley (spin) sectors but leaves the band dispersion unchanged. The Bloch states $u_{n\bm{k}}$ are also unaffected. Consequently, the orbital magnetization remains the same as in the non-interacting case and does not vary with $U$. To modify the band structure and thereby the orbital magnetization, the nearest-neighbor density-density interaction $V$ is required, as shown in Fig.~\ref{fig:kmh}(a). At small $V$, the band gap is located at the $m$ point and increases linearly with $V$. When $V=1.8$, a topological transition occurs, with the global gap shifting to the $\kappa'$ point. The system then evolves into a trivial CDW state as discussed in the main manuscript. This arises because a large nearest-neighbor interaction $V$ favors the occupation of next-nearest-neighbor sites, effectively transforming the honeycomb lattice into a triangular one and driving a transition from a Chern insulator to a trivial insulator.

It is worth mentioning that the nearest-neighbor interaction $V$ does not affect the band energy at the $\kappa$ ($\kappa'$) point.
This follows from the fact that, at $\kappa$ ($\kappa'$), the KMH wave function is localized on the A (B) sublattice, rendering it insensitive to $V$, which couples the A and B sublattices through the Fock term.
For similar reasons, the third–nearest-neighbor interaction alone does not alter the band energy at the $m$ point. However, in realistic situations where interactions of all ranges are present, the band energy will in general be modified at all $k$ points.

The magnetic field is incorporated into the KMH model via the Peierls substitution $t_{i j}\rightarrow t_{i j} \mathrm{e}^{\mathrm{i}\frac{e}{\hbar}\int_{\bm{r}_i}^{\bm{r}_j} \bm{A}\cdot d\bm{l}}$, where $\bm{A}$ is the vector potential of the magnetic field $\bm{B}$. In our calculation, we set $\bm{B}=B\bm{e}_z$ and use the Landau gauge $\bm{A}=-By\bm{e}_x$. To realize a small magnetic field, we introduce one flux quantum into a $12\times12$ unit cell. The resulting energy levels without (left of Fig.~\ref{fig:levels}(b)) and with (right of Fig.~\ref{fig:levels}(b)) the magnetic field are shown. Fig.~\ref{fig:levels} demonstrates that this approach yields energy levels identical to those of the band structure computed on a $12\times12$ $k$-mesh. Connecting the two extremes in Fig.~\ref{fig:levels}(b), we find that one energy level from the valence manifold crosses the band gap and merges into the conduction manifold. This behavior follows from the fact that the first band in the KMH model carries Chern number $C=-1$. According to \cite{Xiao_Berry_2005}, the density of states in phase space is modified by the Berry curvature as $D=(2\pi)^{-d}(1+e/\hbar \bm{B}\cdot \Omega)$, where $d$ is the system dimension, $\bm{B}$ is the magnetic field, and $\Omega$ is the Berry curvature. With a negative Chern number, the number of states hosted by the first band decreases by one.

Because the occupation changes in topological systems under a magnetic field, we compute the grand potential $K=E-\mu N$—which accounts for the total electron number $N$—rather than the total energy $E$. We also verified that the system with one electron less minimizes the grand potential.

\begin{figure}[h]
\centering
\includegraphics[width=0.5\columnwidth]{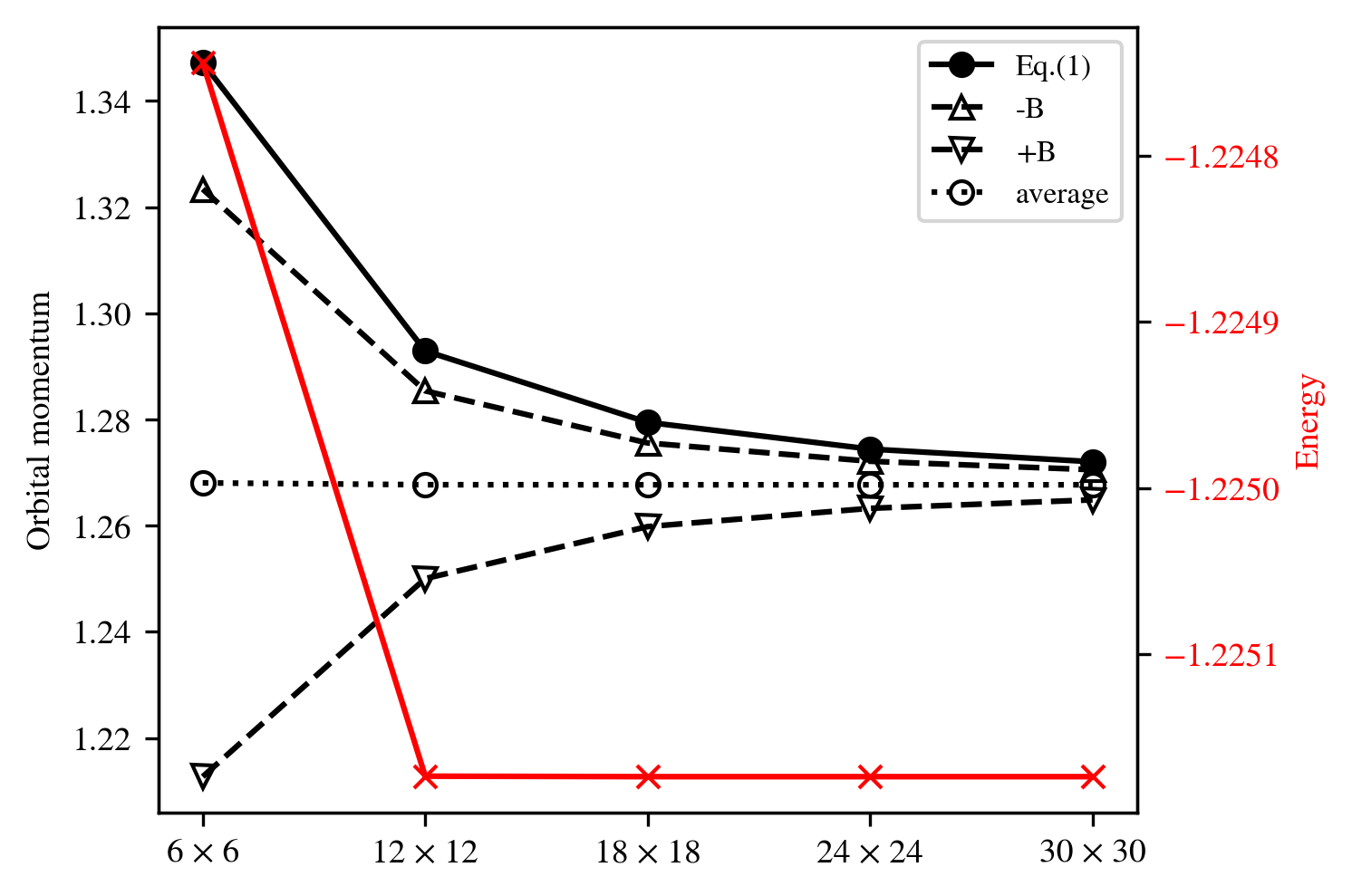}
\caption{Black dots and red crosses are the orbital magnetization and the total energy of the KMH model calculated at different $k$-mesh as listed in the horizontal axis. The triangles are orbital magnetizations calculated by the finite-difference method at different supercell sizes as listed by the horizontal axis. \label{fig:convergence}}
\end{figure}

As we have now employed two methods to calculate the orbital magnetization—one based on Eq.~\eqref{eq:orbmom_hf} and the other from the change of the grand potential—we next examine their numerical stability. We refer to the latter as the finite-difference method.
For the former approach, we tested convergence with respect to the $k$-mesh sampling. As shown by the black dots in Fig.~\ref{fig:convergence}, the orbital magnetization decreases as the Brillouin-zone sampling becomes denser, and eventually converges with further refinement of the $k$-mesh. For comparison, the red crosses display the total energy as a function of $k$-mesh size, which converges much more rapidly than the orbital magnetization.
For the finite-difference method, we considered both positive and negative magnetic fields. In both cases, the results converge as the supercell size increases.
Overall, the two methods yield consistent values of orbital magnetization, confirming the validity of applying Eq.~\eqref{eq:orbmom_hf} to Hartree–Fock systems.

\section{Twist angle dependent orbital magnetization moiré materials}
\begin{figure}[h]
\centering
\includegraphics[width=0.8\columnwidth]{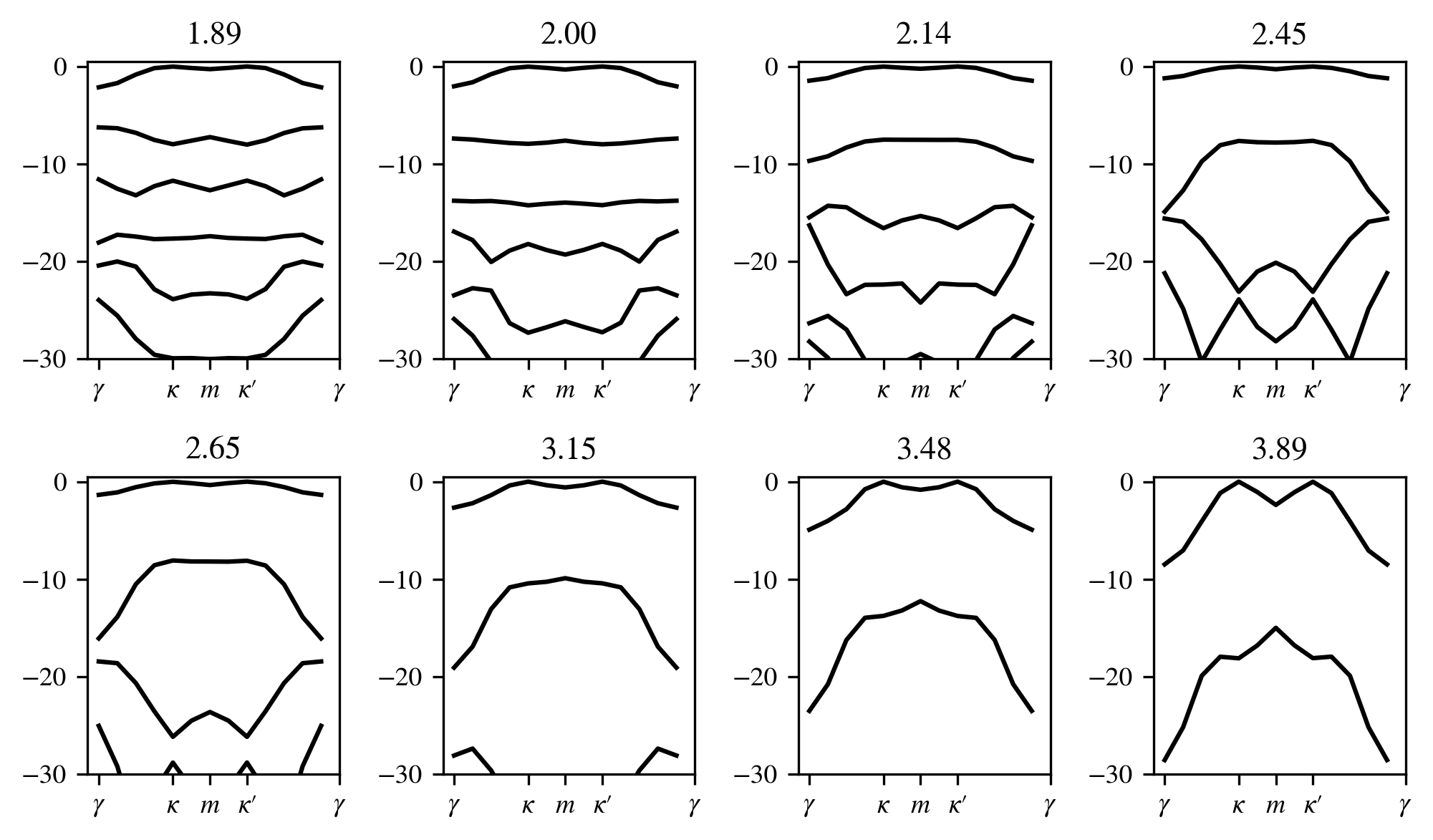}
\caption{Non-interacting band evolution for tMoTe$_2$ at different twist angles. \label{fig:band-nonint}}
\end{figure}

\begin{figure}[h]
\centering
\includegraphics[width=0.8\columnwidth]{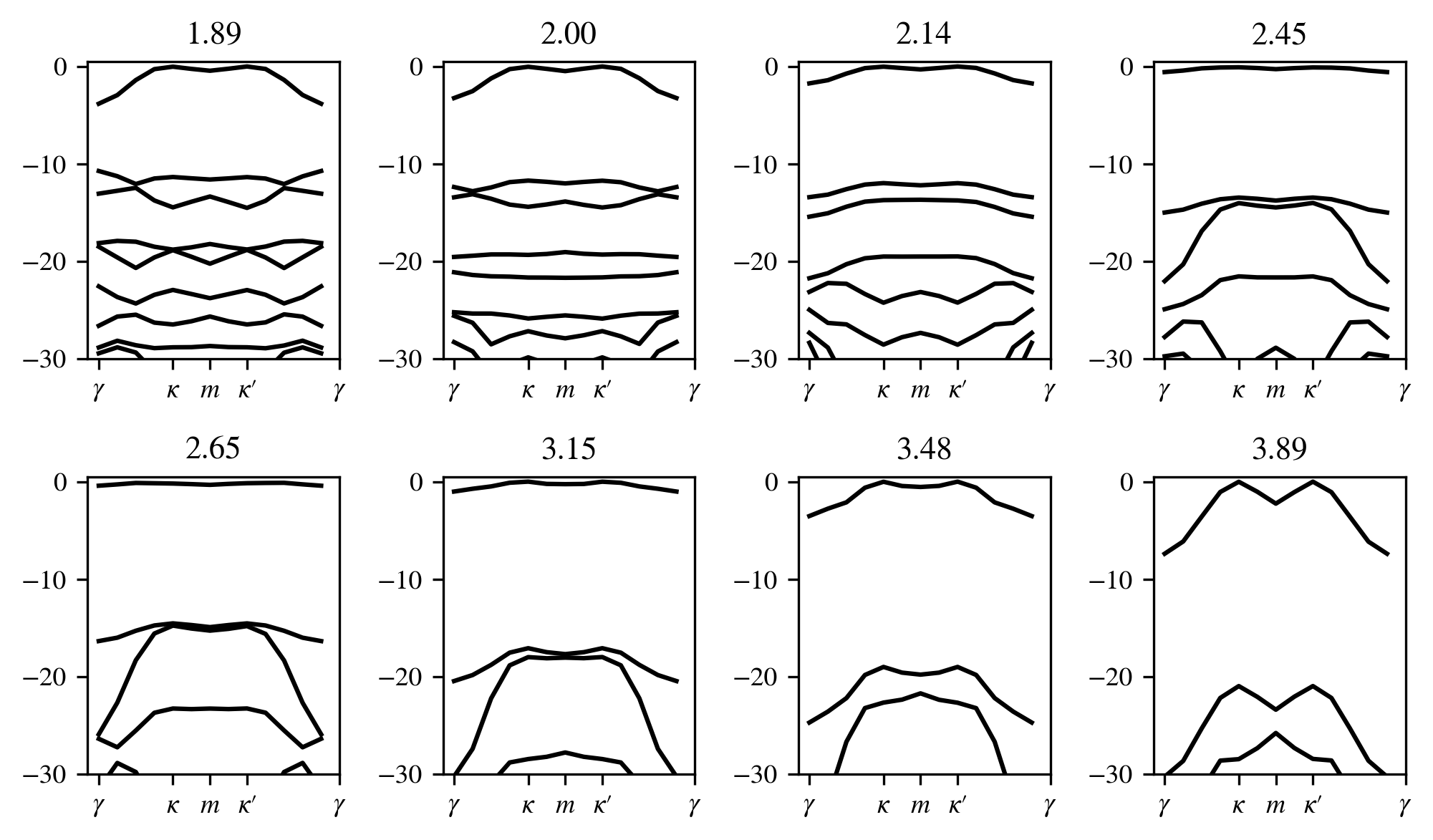}
\caption{Hartree-Fock band evolution for tMoTe$_2$ at different twist angles. \label{fig:band}}
\end{figure}

For each twist angle considered, we construct Wannier functions based on \textit{ab initio} calculations following the procedure in \cite{wang2025higher}. Each Wannier model contains six bands per valley (12 bands in total), and the $k$-mesh is chosen as $12\times12$. The resulting non-interacting band structures are shown in Fig.~\ref{fig:band-nonint}; each band is doubly degenerate since the two valleys are related by time-reversal symmetry.
Hartree–Fock calculations are then performed at filling $\nu=-1$. The dual-gate Coulomb interaction is implemented following Refs.~\cite{chatterjee2020symmetry,liu2024gate}, with a gate–sample distance of 30~nm, interlayer spacing of 7.3~\AA, and dielectric constant $\epsilon=40$. A relatively large dielectric constant is adopted to compensate for the Hartree–Fock tendency to overestimate the gap and to match the experimental values. The resulting Hartree–Fock band structures are shown in Fig.~\ref{fig:band}. The Coulomb interaction lifts the valley degeneracy, polarizing the system in valley (spin) space and opening a large gap between the topmost band and the rest.

\section{Continuum model}
We did not adopt the more commonly used continuum model~\cite{wu2019topological,wang2024fractional,reddy2023fractional} in the main text because the first-shell approximation fails to capture structural relaxations and electronic polarization effects~\cite{zhang2024polarization,jia2024moire}. Continuum models amended with higher shells~\cite{zhang2025twist,zhang2024universal} are cumbersome to implement and have limited transferability across twist angles.

\begin{figure}[h]
\centering
\includegraphics[width=0.5\columnwidth]{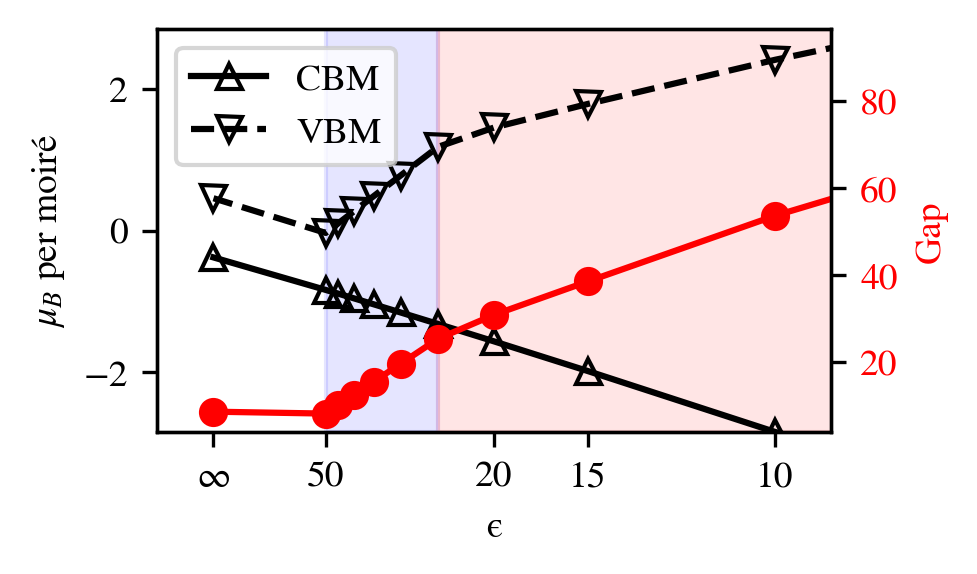}
\caption{Orbital magnetization for VBM and CBM under different dielectric constants. Red dots denote the gap at $\nu=-1$. The values at $\infty$ are calculated at non-interacting limit for a $K$ valley. \label{fig:continuum}}
\end{figure}

Nevertheless, for comparison we also evaluate the orbital magnetization using Eq.~(\ref{eq:orbmom_hf}) within the first-shell continuum model of Ref.~\cite{wu2019topological} and the parameters of Ref.~\cite{wang2024fractional}. We compute the orbital magnetization at various dielectric constants. As shown in Fig.~\ref{fig:continuum}, both the gap (red dots) and the orbital magnetization (triangles) scale linearly with interaction strength (i.e., inversely with $\epsilon$). The orbital magnetization at the CBM (upward triangles) exhibits a smooth linear increase with the increase of interaction strengths, while that at the VBM (downward triangles) shows a distinct kink near $\epsilon=25$, tracking the corresponding kink in the gap (red dots). This feature allows us to distinguish two regions, blue region and red region in Fig.~\ref{fig:continuum}. In blue region (weak interaction), the gap corresponds to exchange splitting between opposite-spin bands. In red region (strong interaction), the spin splitting exceeds the intrinsic gap within a single spin/valley sector, so that the gap is between same-spin bands.
The non-interacting orbital magnetization of $K$ valley is also computed and put at $\epsilon=\infty$ in Fig.~\ref{fig:continuum}). We find that orbital magnetization at VBM (downward triangles) in red region could be connected smoothly to the non-interacting value. This is consistent with our KMH model where  $U$ is large. At $\epsilon=40$ (which locates inside the blue region), the calculated orbital magnetization change is comparable to the experiment \cite{redekop2024direct}.

\section{Covariant derivative}
When calculating differentials in  Eq.~(\ref{eq:orbmom_hf}), we made use of the covariant derivative method as introduced in~\cite{Ceresoli_Orbital_2006}. 
The ground state projector for an effective Hamiltonian $H_{\bm{k}}$ is defined as 
\begin{equation}
    P_{\bm{k}} = \sum_{n\in \textrm{occ}}|u_{n\bm{k}}\rangle \langle u_{n\bm{k}}|.
\end{equation}
where $u_{n\bm{k}}$ is the eigenstates of $H_{\bm{k}}$.
$Q_{\bm{k}}=1-P_{\bm{k}}$ is the complementary projector projecting to unoccupied space. 
According to~\cite{Ceresoli_Orbital_2006}, the orbital magnetization can be written as
\begin{equation}
\bm{M} = \frac{e}{2\hbar}i\sum_{n\bm{k}}
\bigg\langle \tilde{\partial}_{\bm{k}} u_{n\bm{k}}\bigg|
\times[2\mu-\hat{H}_0(\bm{k})-\epsilon_n(\bm{k})]
\bigg| \tilde{\partial}_{\bm{k}} u_{n\bm{k}} \bigg\rangle,
\end{equation}
where 
\begin{equation}
|\tilde{\partial}_{\bm{k}} u_{n\bm{k}}\rangle = Q_{\bm{k}}| u_{n\bm{k}} \rangle
\end{equation}
is the covariant derivative. 
By definition, the covariant derivative has the property that $\langle u_{n'\bm{k}} | \tilde{\partial}_{\bm{k}} u_{n\bm{k}}\rangle = 0$. 

We can construct a "dual" state $\tilde{u}_{n\bm{k+q}}$ that is a linear combination of $u_{n\bm{k+q}}$ (where $n$ goes over occupied bands) and satisfies the above condition as $\langle u_{n'\bm{k}} | \tilde{u}_{n\bm{k+q}}\rangle=\delta_{n'n}$ simultaneously. This gives
\begin{equation}
|\tilde{u}_{n\bm{k+q}}\rangle=\sum_{n'\in \textrm{occ}} (S_{\bm{k,k+q}}^{-1})_{n'n} |u_{n'\bm{k+q}}
\end{equation}
with
\begin{equation}
    (S_{\bm{k,k+q}}^{-1})_{nn'}=\langle u_{n\bm{k}}| u_{n'\bm{k+q}}\rangle.
\end{equation}

The covariant derivative is calculated by the finite-difference of the "dual" state
\begin{equation}
    |\tilde{\partial}_{\bm{k}} u_{n\bm{k}}\rangle=\frac{1}{2}(|\tilde{u}_{n\bm{k+q}}\rangle - |\tilde{u}_{n\bm{k-q}}\rangle).
\end{equation}

\end{document}